\documentclass[aps,prd,superscriptaddress,tightelines,nofootinbib]{revtex4}
\usepackage{amssymb,amsmath}
\usepackage{latexsym,amssymb,amsmath,amsfonts,amssymb,txfonts,pxfonts,wasysym,float}
\usepackage[dvips]{graphics,color}
\usepackage{epsfig}
\usepackage{lmodern}
\usepackage{subfig}
\usepackage{graphicx}

\begin{document}

\title{Galactic Centre as an efficient source of cosmic rays}

\author{Rita. C. Anjos}
\email{ritacassia@ufpr.br}

\affiliation{Department of Astronomy, Harvard University, 60 Garden Street, Cambridge, MA 02138, USA}
\affiliation{Departamento de Engenharias e Ci\^encias Exatas,
  Universidade Federal do Paran\'a (UFPR),\\ 
Pioneiro, 2153, 85950-000 Palotina, PR, Brazil.}

\author{Fernando Catalani}
\email{fcatalani@usp.br}

\affiliation{Escola de Engenharia de Lorena, Universidade de S\~ao Paulo, \'Area I - Estrada Municipal do Campinho, S/N, 12602-810 Lorena, SP, Brazil.}

\begin{abstract}  

After the discovery of Fermi Bubbles and the excess of gamma-ray emission, the Galactic Centre has received increasing attention with the aim to understand its role in the origin and acceleration of primary cosmic rays (CRs). Based on a diffusion/re-acceleration model, we use the GALPROP software to solve the diffusion equation for the cosmic rays and compare the results with the CRs spectrum, motivated by relationship between several energetic sources at the Galactic Centre and the generated diffuse GeV-TeV gamma-rays. We calculate the cosmic ray distribution, gamma-ray flux from Galactic Centre and explore its contribution on the spectrum and chemical composition of cosmic rays observed at Earth. We also discuss the effects of nuclei interactions with different interstellar gas models. 

\end{abstract}

\maketitle

\section{INTRODUCTION}

The origin of cosmic rays with energies up to $10^{15}$ eV are commonly attributed to supernova remnants (SNR) in our Galaxy \cite{reynolds}.  However, with a supermassive black hole, millisecond pulsars, transient radio and X-ray sources located around the Galactic Centre (GC), it has become an excellent scenario for the study of astrophysical phenomena, indicating a correlation between past activities in the GC region and observational evidence \cite{cheng, fujita, guepina, lacki}. 
Historically, the models for the sources radial distributions consider that the density vanishes at the GC. Since our Galaxy is a spiral one, the population of SNR is distributed over the entire galactic disk, leaving the parametrizations for the density of sources zero at $r = 0$ \cite{eric, green}. However, recent observations with remarkable sensitivity were able to detect activities in the GC \cite{fermi, hess}, with the discoveries of the 511 keV emission line and Fermi Bubbles (FB) \cite{cheng, fujita, guepina, lacki}. 

Much effort has been devoted to the study and observation of the GC since it is one of the most appealing regions observed from radio to gamma ray. The observations of the GC confirm the presence of a Supermassive black hole (Sagittarius A*) with an intense activity in the past. Outstanding recent discoveries indicate that Sgr A* can be a source of PeV galactic cosmic rays \cite{hess}, that could have been accelerated from relativistic outflows. Based on this scenario, Cheng et al \cite{cheng} propose that Fermi Bubbles can act as a acceleration site for some of the CR particles originally accelerated in SNR located at galactic disk. In their model, the origin of the Bubbles is explained by the periodic capture and tidal disruption of stars by the Black Hole at the Galactic Centre. These gargantuan events create periodic energy releases in the Halo in the form of multiple shock waves, with larger length scales and lifetime than those found in SNR, allowing the stochastic acceleration of particles inside the Bubbles at energies higher than $10^{15}$ eV.
 Very recently, the Atacama Large Millimeter/sub-millimeter Array (ALMA) and measurements of stellar orbits around Galactic Centre reported indications of new candidates in the central region of our Galaxy besides Sgr A*. These discoveries can indicate much more activity in the past which could explain the TeV $\gamma$-ray observations \cite{take,naoz}.

The CRs interact with the molecular gas via nuclear reactions in the galaxy and generate diffuse GeV-TeV gamma-rays. Many scenarios have been proposed to explain the GC excess of gamma ray emission at GeV energies up to $\sim$ 20$^{\circ}$ from the GC generated by interactions in the galaxy. Detailed discussions of these interpretations can be found in \cite{fujita, cheng, crocker, eric, petro}. In addition, the GC excess energy spectrum is also consistent with a possible scenario of gamma rays emitted from a galactic halo of dark matter \cite{calore, tansu}. However, no indication of the flux from Milky Way dwarf galaxies, expected to be dark matter dominated, has been detected \cite{louis}. 

Hadronic and leptonic models are two distinct interpretations suggested to explain the gamma-ray production from activities related to the GC \cite{cheng, soebur,lacki}. In the hadronic models, gamma rays are produced by decay of neutral pions produced in inelastic collisions between relativistic protons and the thermal nuclei. On the other hand, in the leptonic reactions, gamma rays are generated by inverse-Compton scattering of background interstellar radiation by cosmic-ray electrons and positrons \cite{soebur,fujita}. In \cite{nicola} is suggested a connection between leptonic and hadronic anomalies with the objective of modeling the diffusive gamma rays emission in other parts of the Galaxy, taking into account old SNR contribution to the GeV-TeV spectrum.

In this paper, based on recent cosmic rays and gamma ray observations \cite{crocker, fang}, we calculate the contributions of cosmic rays nuclei spectra and gamma rays spectra with a numerical implementation of models of the GC region \cite{jaupart, guo, abra, acero}. We extend the treatment of the case given in \cite{jaupart} using GALPROP \cite{strong}. We consider two classes of CRs sources represented by a source at the centre of the Galaxy (Model GC) and sources modeled by a spherical power-law density profile around Galactic Centre (Model FB) \cite{kataoka}. This paper is organized as follows: in Section II we describe our CRs propagation calculations. In particular we detail the numerical implementation of our model and the methodology adopted for the determination of the key parameters. In Section III we present our results for the models and discuss the gamma-rays from the GC and the conclusions are summarized in Section IV.

\section{DESCRIPTION OF THE MODEL}

We use the numerical code GALPROP v56 to simulate the distribution of CRs in the Galaxy \cite{strong}. GALPROP solves the transport equation for a given source distribution and boundary conditions using the second-order Crank-Nicolson method. The equation includes energy losses, nuclear fragmentation and decay, source distributions, convection (galactic wind), diffusive and re-acceleration processes. We adopt the diffusion/re-acceleration model, which describes well the secondary-to-primary ratios and the effects of convection. A detailed description of the GALPROP model is given in \cite{strong, moskalenko, strong1, moskalenko1} and Web Site \cite{web}. The GALPROP software has been used in previous studies of Galactic Centre \cite{eric,profumo,fermilat1} and in this work we extended the code in order to analyze the contribution of CRs sources at the centre of the Galaxy.

Our simulations assume cylindrical geometry in the galaxy with a diffusion halo with radius $R = 20$ kpc and halo height $z = 10$ kpc with $z = 0$ located at the galactic plane, which are suggested by previous studies \cite{luka}. Since the Fermi Bubbles (FB) extend up to 9 kpc north and south from the GC and the halo size has influence on the CRs fluxes, the variations related to the geometry are important to the final cosmic rays and gamma-rays spectra. Using a smaller halo height for the Galactic Centre region as observed by the WMAP haze \cite{biermann}, a very large injection power will be needed to describe the CRs data \cite{jaupart}. Taking into account the diffusion and re-acceleration model, the transport equation for cosmic-rays nuclei in the Galaxy can be written as \cite{web}:
 \begin{equation}
 \frac{\partial N}{\partial t} = Q(\textbf{x},p) + \vec{\nabla}.(D_{xx}\vec{\nabla}N - \vec{v} N) + \frac{\partial}  {\partial p}p^{2}D_{pp}\frac{\partial}{\partial p}\frac{1}{p^2}N - \frac{\partial}{\partial p}(\dot{p}N -\frac{p}{3}(\vec{\nabla}.\vec{v})N) - \sum_{i=1}^{2}\frac{N}{\tau_{i}} 
 \label{eq:transport}
\end{equation}
where $N(x,p)$ is the cosmic rays density per unit of total particle momentum and $Q$ is a cosmic rays injection source term including secondary production of cosmic-rays. In our model, $D_{xx}$ is the spatial diffusion coefficient, $\vec{v}$ is the convection velocity, the time scales for fragmentation and radioactive decay are $\tau_{1}$ and $\tau_{2}$, respectively \cite{strong}.

The diffusion coefficient is assumed as a scalar function, homogeneous and isotropic throughout the Galaxy, depending on the particle rigidity via a power law with an index $\delta$: $D_{xx} = \beta D_{0}(\rho/\rho_0)^{\delta}$, where $\rho_0 = 3$ GV, $D_0$ is a diffusion constant, $\beta$ is the velocity in unit of light speed. Cosmic-ray propagation models commonly assume an isotropic diffusion coefficient to account for the random deflection of cosmic rays by the interstellar magnetic field. In our model we consider a spatial diffusion inversely proportional to the magnetic field turbulent component $D(\rho,z) = D_{xx}\exp (\frac{\left |z \right |}{z_t})$, where $z_t$ is a characteristic scale (halo size $\sim 3z_t$) \cite{carmelo}. However, an inhomogeneous anisotropic diffusion coefficient is essential in future models to interpret recent measurements of large scale anisotropy of TeV cosmic rays and gamma-ray diffuse emissions \cite{wel}.

The convection velocity $\vec{v}(z) =  dv/dz \times z$ (in z-direction only) is assumed to increase linearly with distance $z$ from the plane to halo. The cosmic rays propagation is diffusive-convective in one zone $\left |z \right | > 1$ kpc \cite{ptuskin} and both galactic wind and energy losses are important at low energies, up to $\sim 1$ GeV \cite{thoudam}. The re-acceleration is determined by a momentum diffusion coefficient $D_{pp}$, which is related to the spacial coefficient $D_{xx}$ via $D_{pp} \propto p^{2} v_{a}^{2}/D_{xx}$, where $v_{a}$ is the Alfv\'en speed, $\dot{p}\equiv dp/dt$ is the momentum loss rate and the Kolmogorov spectrum of turbulence is considered. The parameters of the diffusion coefficient and Alfv\'en speed are determined by $B/C$ ratio \cite{strong, carmelo}.

Our model of the galactic magnetic field has a cylindrical symmetry and is chosen to provide a description of the synchrotron emission of the Galaxy: $ B(r,z) = B_{0}e^{(R_{\odot}-r)/R_{B}}e^{-\left |z \right |/z_{B}}$ \cite{strong1}, where $R_{\odot}= 8.5$ kpc is the Solar Radius, $B_{0} = 5\mu$ G and $R_{B} = 6$ kpc and $z_{B} = 2$ kpc are the radial and vertical scale-lengths \cite{strong}. This model framework only considers the random magnetic field distribution, since it is associated to diffusive properties of cosmic rays \cite{carmelo,wel}. However, the strong magnetic fields are insufficiently constrained near the galactic centre \cite{profumo} and are beyond the scope of the present model.

The source function $Q(\textbf{x},p)$ is described as $n(\textbf{x})q(p)$, where $n(\textbf{x})$ is the spatial distribution and $q(p)$ is the injection energy spectrum of cosmic rays. We adopt a differential injection spectrum which follows a power law in momentum $dq(p)/dp \propto p^{-\alpha}$, with index $\alpha = 2.4$ shown in \cite{jaupart} for proton. For the spatial distribution we consider the contribution of a continuous source of energetic particles at the center of the Galaxy. We assume a continuous emission for a period of $10^{7}$ yr. In our models, the CRs injection power is in the range $\sim$ 2 - 4 $\times 10^{41}$ erg.s$^{-1}$, power needed for the luminosity of CRs in our galaxy.

The High Altitude Water Cherenkov (HAWC) gamma-ray observatory reported upper limits above $1$ TeV in the Northern Fermi Bubble region \cite{hawc}. In the hadronic models, the upper limit on the integral flux of GeV-TeV gamma-rays of a given source can lead to the upper limit on the total CRs luminosity \cite{vitor, anjos}. Motivated by this correlation, to model the spatial distribution of the sources at the FB, we adopt the distribution of the galactic gas halo in which FB expands \cite{kataoka, zhang}: $n(r) \propto [1+ (\frac{r}{r_c})^{2}]^{-3\xi/2}$, where $r$ is the distance to the GC, $r_{c} = 0.35$ kpc is the core radius and $\xi = 0.71$. On the other hand, the Model GC injects CRs following a $\delta$-function centered at the centre of the Galaxy.

\section{RESULTS AND DISCUSSION}

\begin{figure}[h!]
  \centering
    \subfloat[Proton]{\includegraphics[angle=0,width=0.5\textwidth]{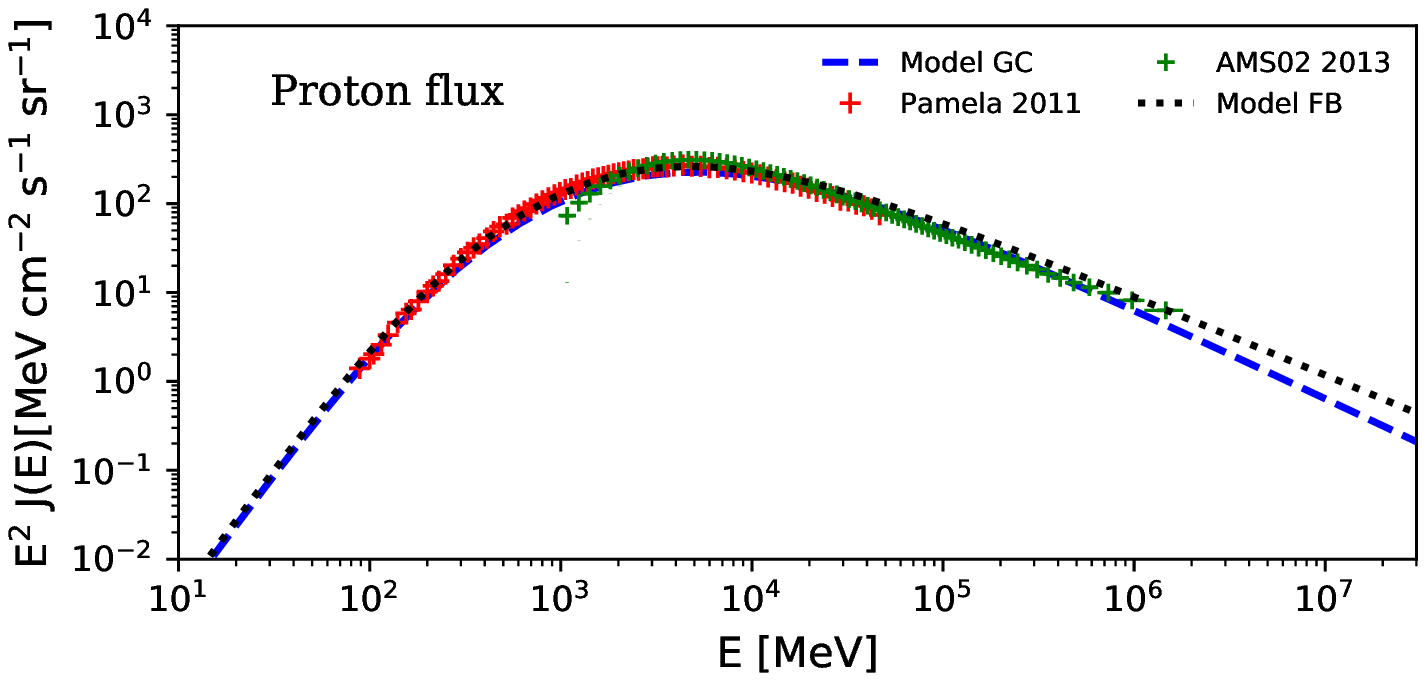}}
    \subfloat[Helium]{\includegraphics[angle=0,width=0.5\textwidth]{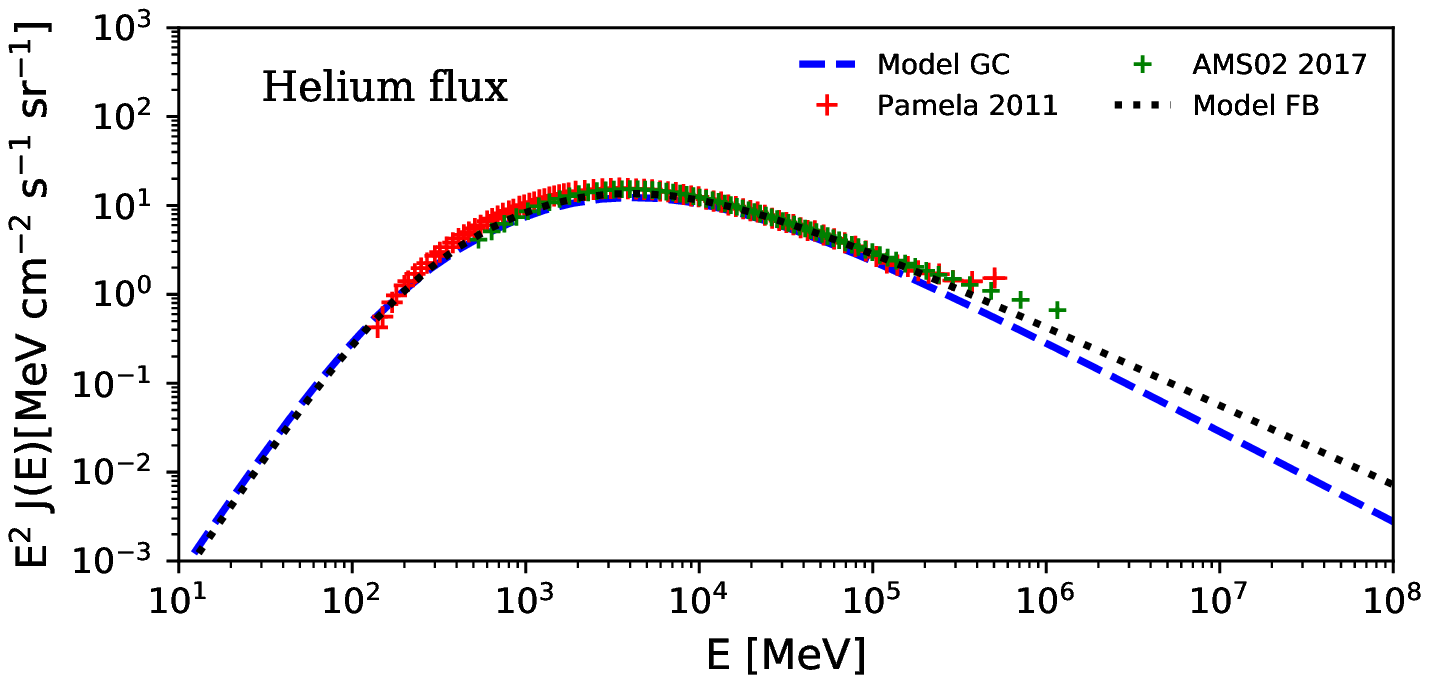}}\\
    \subfloat[Carbon]{\includegraphics[angle=0,width=0.5\textwidth]{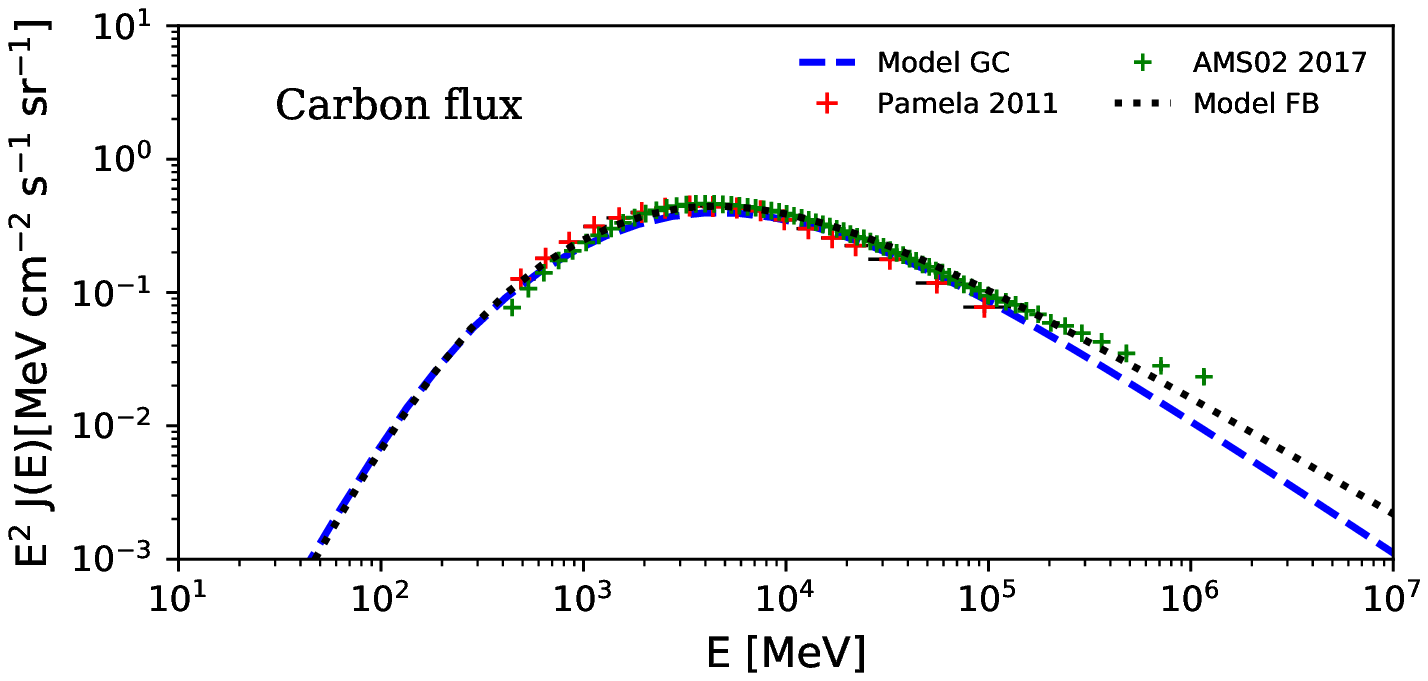}}
    \subfloat[Oxygen]{\includegraphics[angle=0,width=0.5\textwidth]{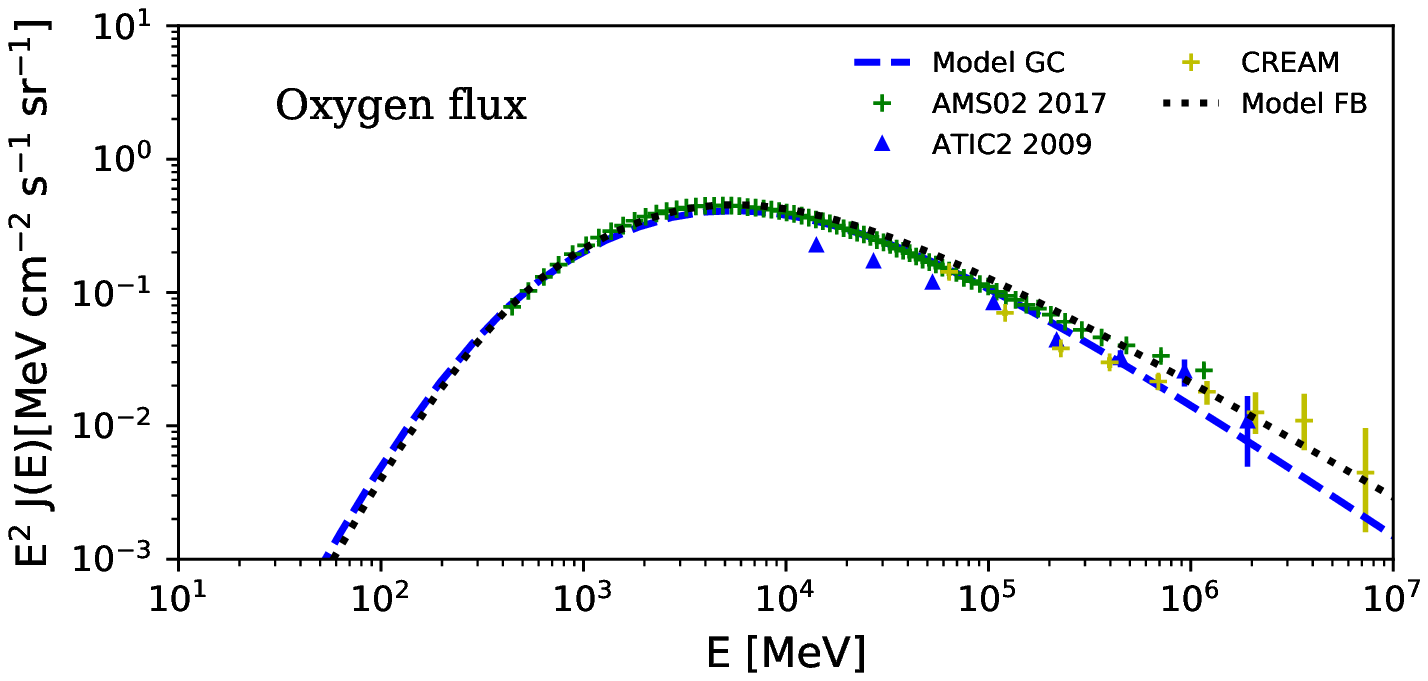}}\\
    \subfloat[Silicon]{\includegraphics[angle=0,width=0.5\textwidth]{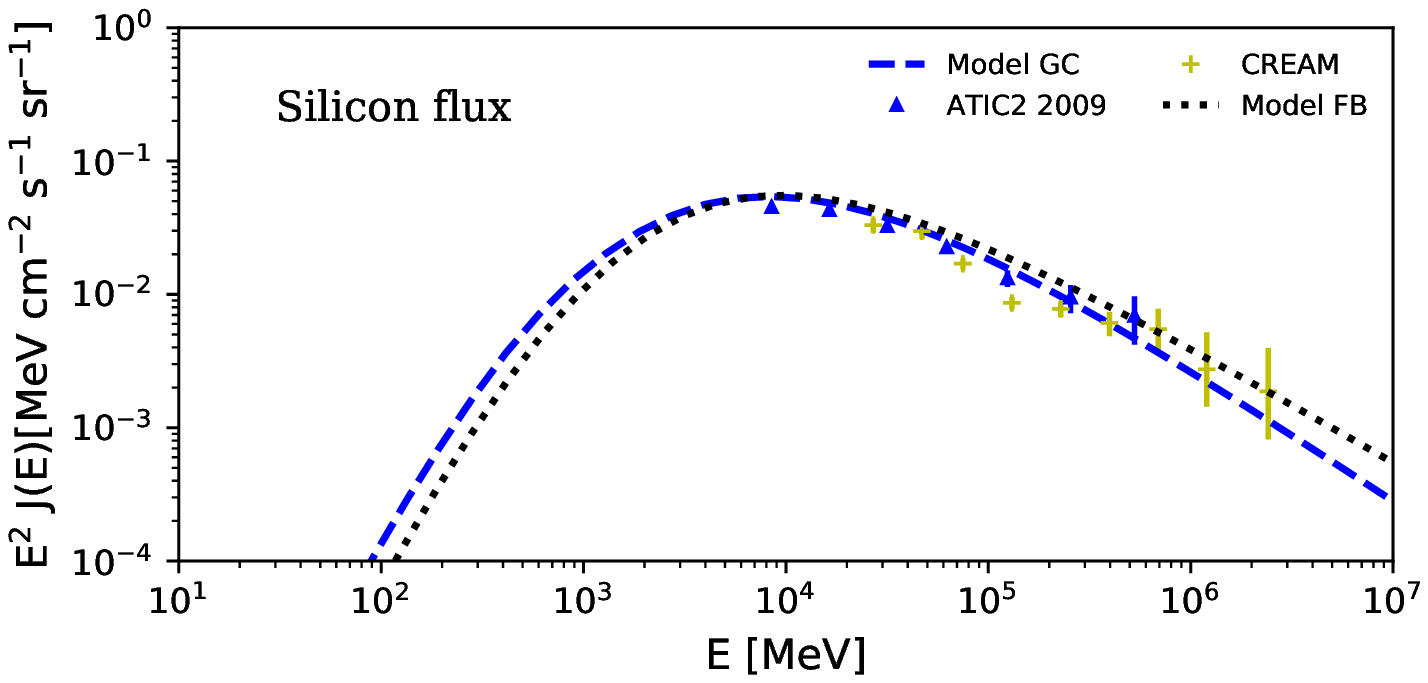}}
    \subfloat[Iron]{\includegraphics[angle=0,width=0.5\textwidth]{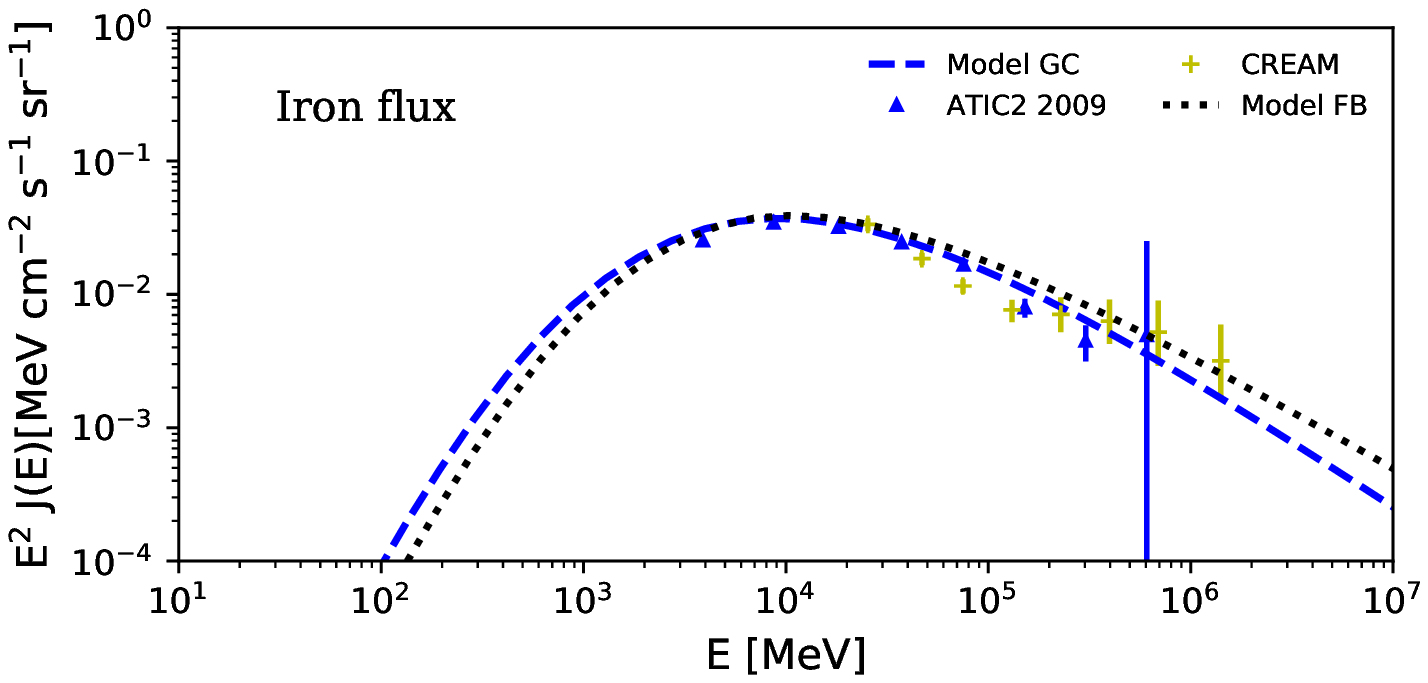}}\\
    \caption{Energy spectra for six elements: proton, helium, carbon, oxygen, silicon and iron. The spectra are multiplied by $E^{2}$. The models calculations are shown in comparison with the data using different modulation parameter values for elements, respectively: $\Phi = 0.30,0.30,0.30,0.40,1.0$ GV for Model GC and $\Phi = 0.30,0.35,0.35,0.50,1.5$ GV for Model FB. The data are extracted from the CRDB database \cite{crdb}.}
    \label{spectra} 
\end{figure}

\begin{figure}[h!]
  \centering
   \subfloat[Boron to Carbon ratio]{\includegraphics[angle=0,width=0.5\textwidth]{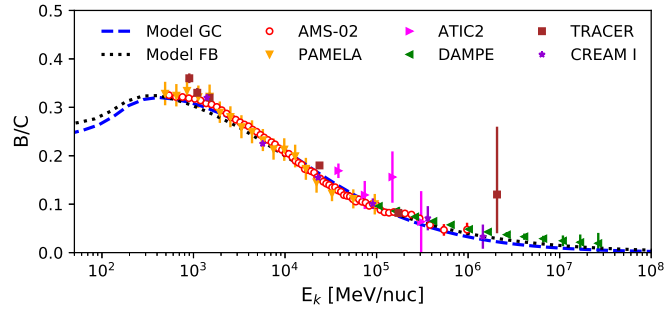}}\\
   \subfloat[Boron to Carbon ratio at high energy ($K \geq 10^{5}$ MeV)]{\includegraphics[angle=0,width=0.5\textwidth]{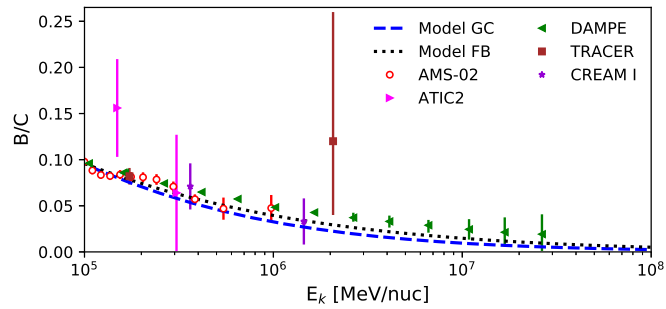}}
    \caption{ $B/C$ ratio as function of kinetic energy per nucleon for the models GC and FB.}
    \label{ratio}
\end{figure}

Considering that the purpose of this paper is to investigate the interplay of galactic cosmic-rays and GC gamma-rays, we compare the measured data at low energies with the energy spectra of Model GC and Model FB for different elements (proton, helium, carbon, oxygen, silicon and iron) in figure \ref{spectra}. The parameter values that reproduce the measured data at low energies are listed in table \ref{tab:1}. The halo half-thickness $z = 10$ kpc was determined by fitting the $B/C$ ratio at high energy \cite{genolini}, see figure \ref{ratio}a. The parameters of the Model GC are in agreement with those in \cite{jaupart}. It shows that the CRs flux for lower energies is comparable to that from the SNR and suggests that one single powerful source at the GC centre could provide enough CRs that are detected on Earth. Small variations on the values of re-acceleration and convection does not change significantly our results. As is clear from fig. \ref{spectra}, the model FB reproduces well the observed data up to $\sim$ TeV. This indicates the effect of halo size on the resulting energy spectrum of CRs \cite{strong}. 

\begin{figure}[h!]
  \centering
  \includegraphics[angle=0,width=0.5\textwidth]{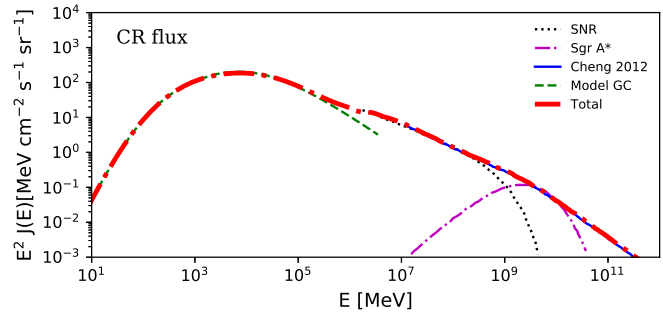}
    \caption{Contribution of GC - SNR - SgR A* to the proton spectrum. The red line represents the observations and total contribution. The model GC is the CRs flux calculated in this paper. The Cheng model 2012 \cite{cheng} describes the re-acceleration of the CRs in the Fermi Bubbles producing CRs beyond the knee. These CRs are produced by SNR in the galactic disk. The lines black and magenta represent the contribution of CRs from SNR and spectrum of CRs injected by Sgr A*, respectively \cite{fujita}.}
    \label{spectrum}
\end{figure}

The CRs injected by sources located at the GC can contribute to the spectrum up to knee \cite{cheng}. Consequently, the CRs spectrum is a combination of the SNR contribution from the galactic disk and the CRs acceleration in the FB. Similarly, in \cite{fujita}, based on a diffusion-halo model, Sagittarius A* can contribute as a Pevatron to the galactic CRs near the knee. Our model GC added to the SNR and SgR A* contributions to the proton spectrum is shown in figure \ref{spectrum}. The total proton spectrum is consistent with the measured data and shows that the contribution of the GC can explain the observed primary CRs from GeV up to PeV.     

The measurements of the secondary-to-primary ratio is a tool to understand cosmic rays propagation in the Galaxy. In particular, boron to carbon flux ratio measures the average quantity of interstellar matter traversed by cosmic rays. While boron nuclei are produced by the interactions of heavy nuclei, carbon nuclei are principally produced in the sources. In figure \ref{ratio}, both propagation models provide good fit for $B/C$ ratio as a function of kinetic energy per nucleon. This figure combines data from TRACER \cite{tracer}, PAMELA \cite{pamela}, ATIC2 \cite{atic2}, CREAM \cite{cream}, AMS-02 \cite{ams2} and DAMPE \cite{dampe}, with their statistical uncertainties. The spectra was modulated to 300 MV, appropriate to these data. The Solar modulation is important below a few GeV. At the highest energies ($K \geq 10^{5}$ MeV), all results are based on few events indicating that the primary cosmic rays suffer less spallation and the correction for atmospheric production of boron may become considerable \cite{cowsik,ober}. The agreement with the data at high energy is less accurate, which may be improved considering SNR also as sources and CRs acceleration at the same time.

\begin{figure}[h!]
  \centering
    \subfloat[]{\includegraphics[angle=0,width=0.5\textwidth]{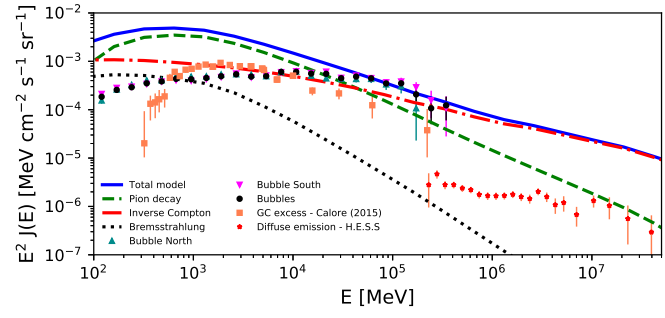}}\\
    \subfloat[]{\includegraphics[angle=0,width=0.5\textwidth]{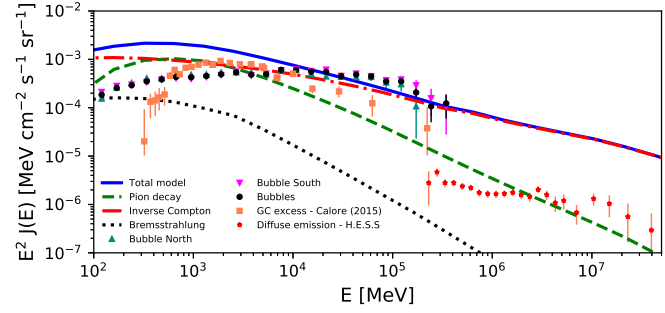}}
    \caption{Inner Galaxy spectra of diffuse gamma-ray emission components. Model GC with a) gas distribution \cite{fermilat} and b) new 2D gas distribution \cite{troy}, respectively. The data are extracted from the CRDB database \cite{crdb}.}
    \label{gamma}
\end{figure}

In figure \ref{gamma}-(a,b) we show the gamma spectrum from inverse Compton scattering, pion decay and bremsstrahlung that we have calculated using the gas distribution adopted by the Fermi-LAT collaboration \cite{fermilat} - \ref{gamma}a and by the new 2D gas distribution galactic ensemble component \cite{troy} - \ref{gamma}b. GALPROP produces a projected map of the gamma ray flux as a product of the cosmic-ray protons propagation taking into account the gas model.

The interstellar gas consists essentially of hydrogen and helium, while heavier elements represent a minor fraction of the total gas mass. The figure \ref{gamma}-(a,b) shows that dominant processes for gamma ray production are inverse Compton scattering and decay of neutral pions. Also, the figure \ref{gamma}b shows that at high energy $\sim$ 1-10 TeV the observations from H.E.S.S may reflect the production from hadronic models, which could be indicating the existence of a PeV proton accelerator around GC \cite{hess}.
 
\begin{figure}[h!]
  \centering
    \subfloat[Bremss]{\includegraphics[angle=0,width=0.32\textwidth]{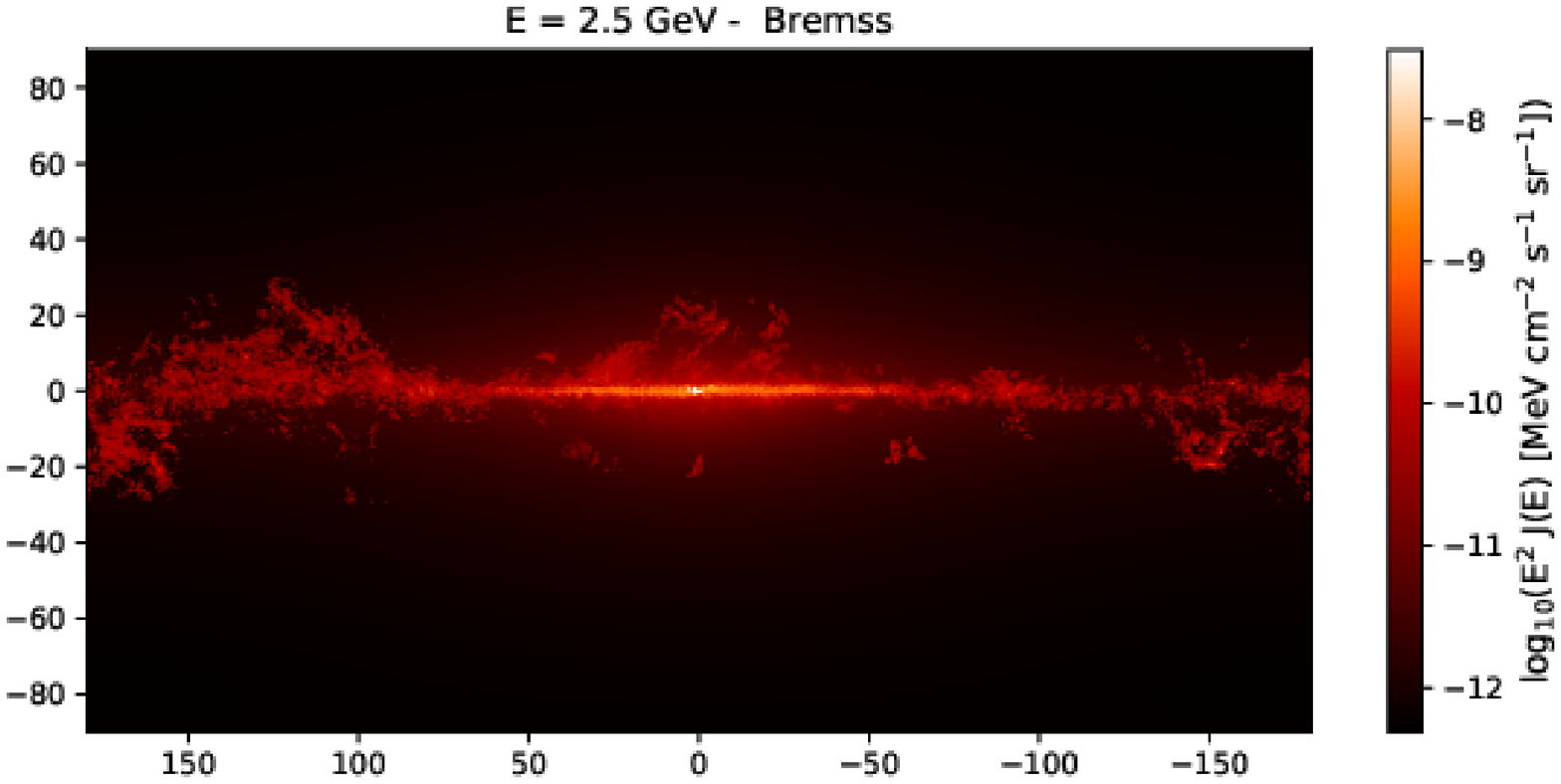}}
    \subfloat[Bremss]{\includegraphics[angle=0,width=0.32\textwidth]{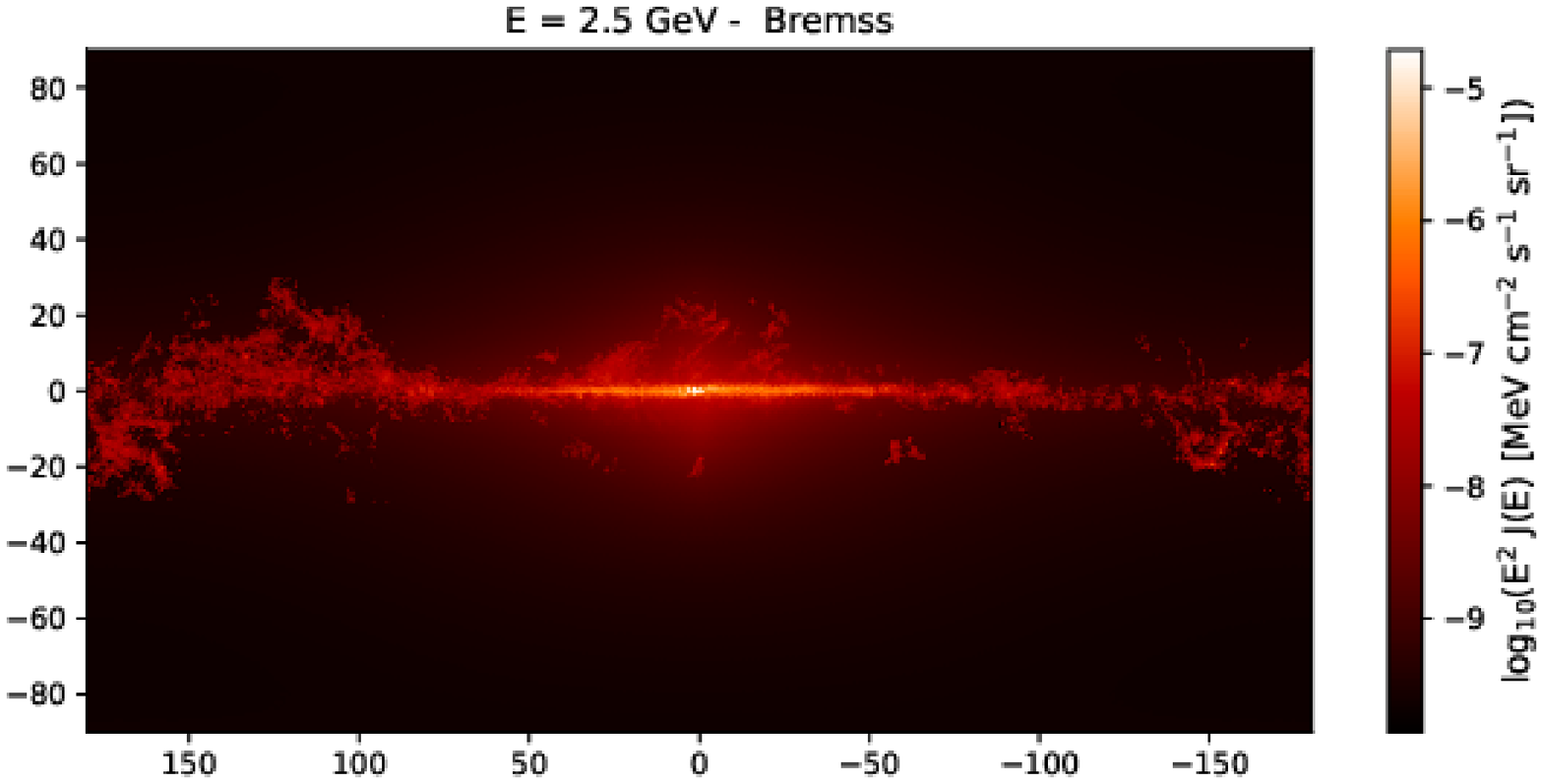}}
    \subfloat[Fractional residual - bremsstrahlung intensity]{\includegraphics[angle=0,width=0.31\textwidth]{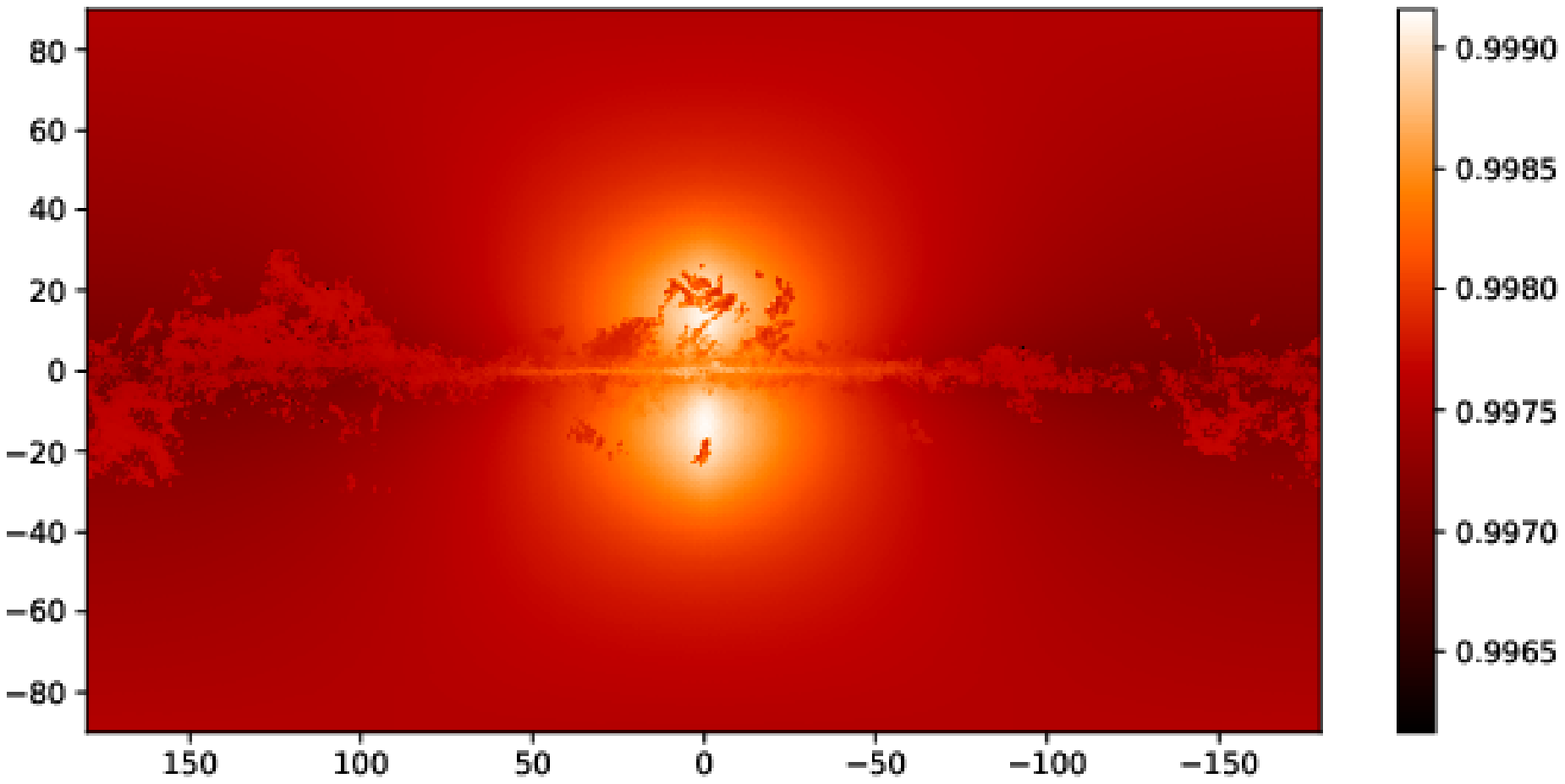}}\\
    \subfloat[Inverse Compton]{\includegraphics[angle=0,width=0.32\textwidth]{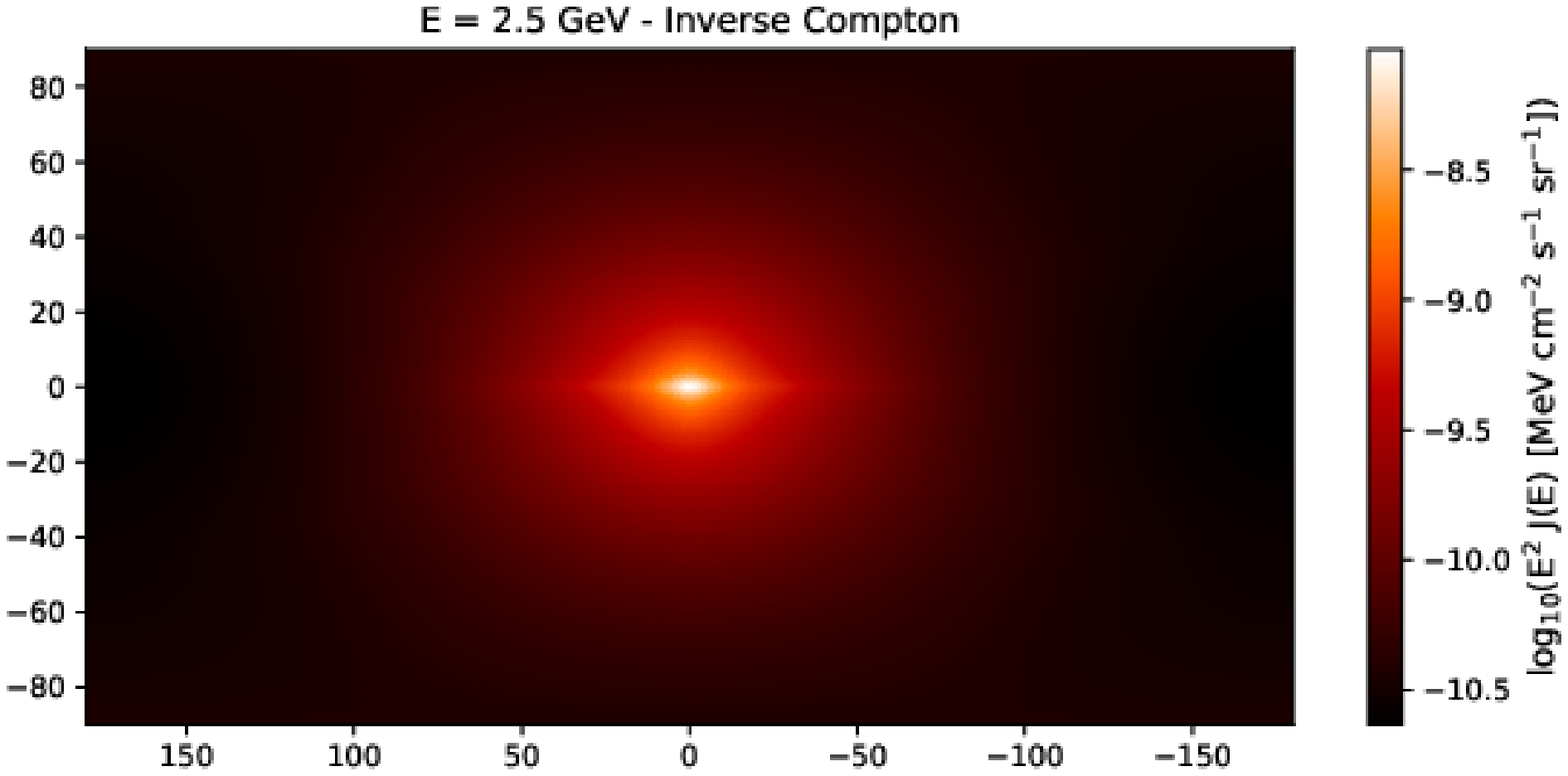}}
    \subfloat[Inverse Compton]{\includegraphics[angle=0,width=0.32\textwidth]{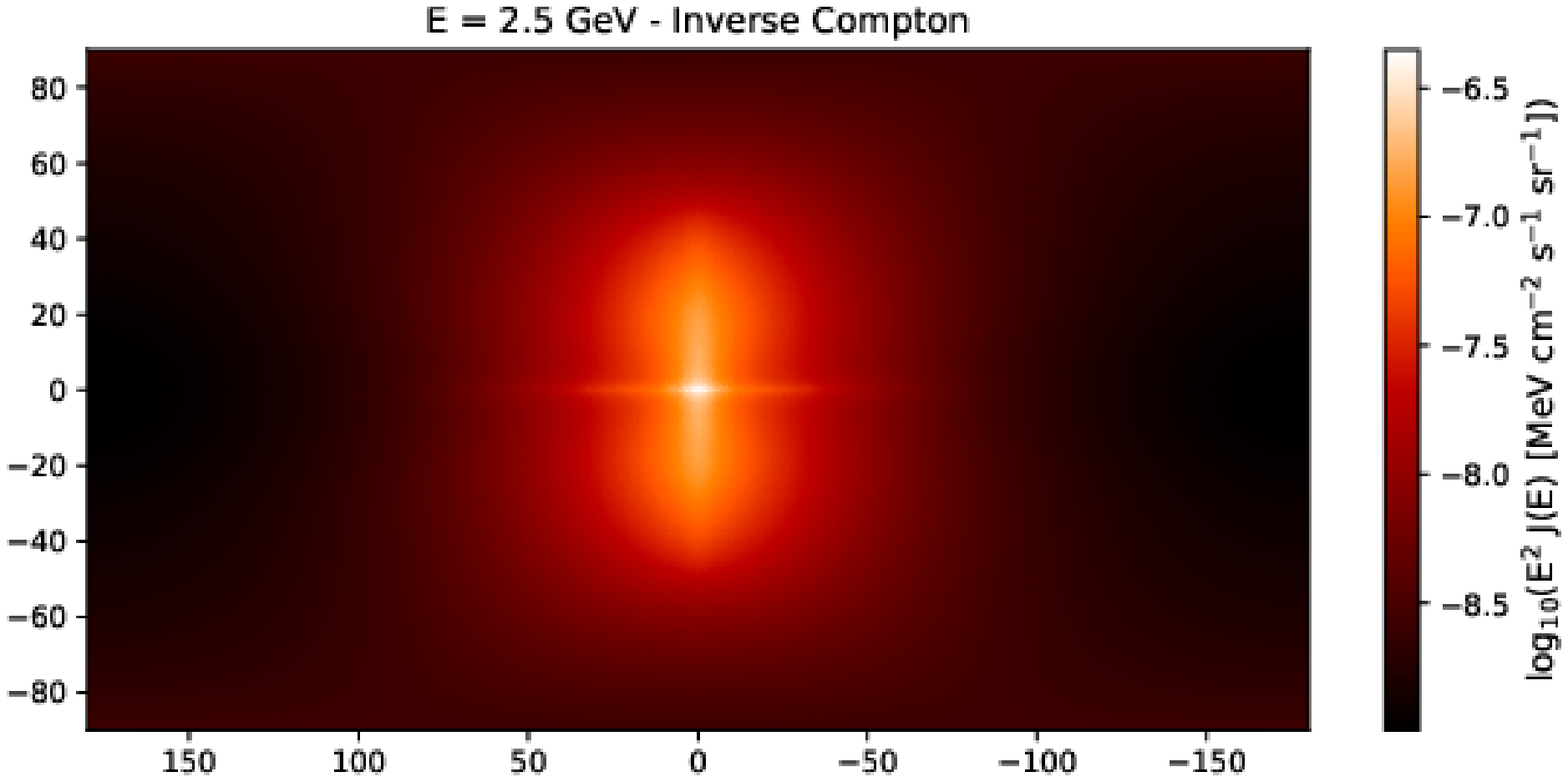}}
    \subfloat[Fractional residual - IC intensity]{\includegraphics[angle=0,width=0.31\textwidth]{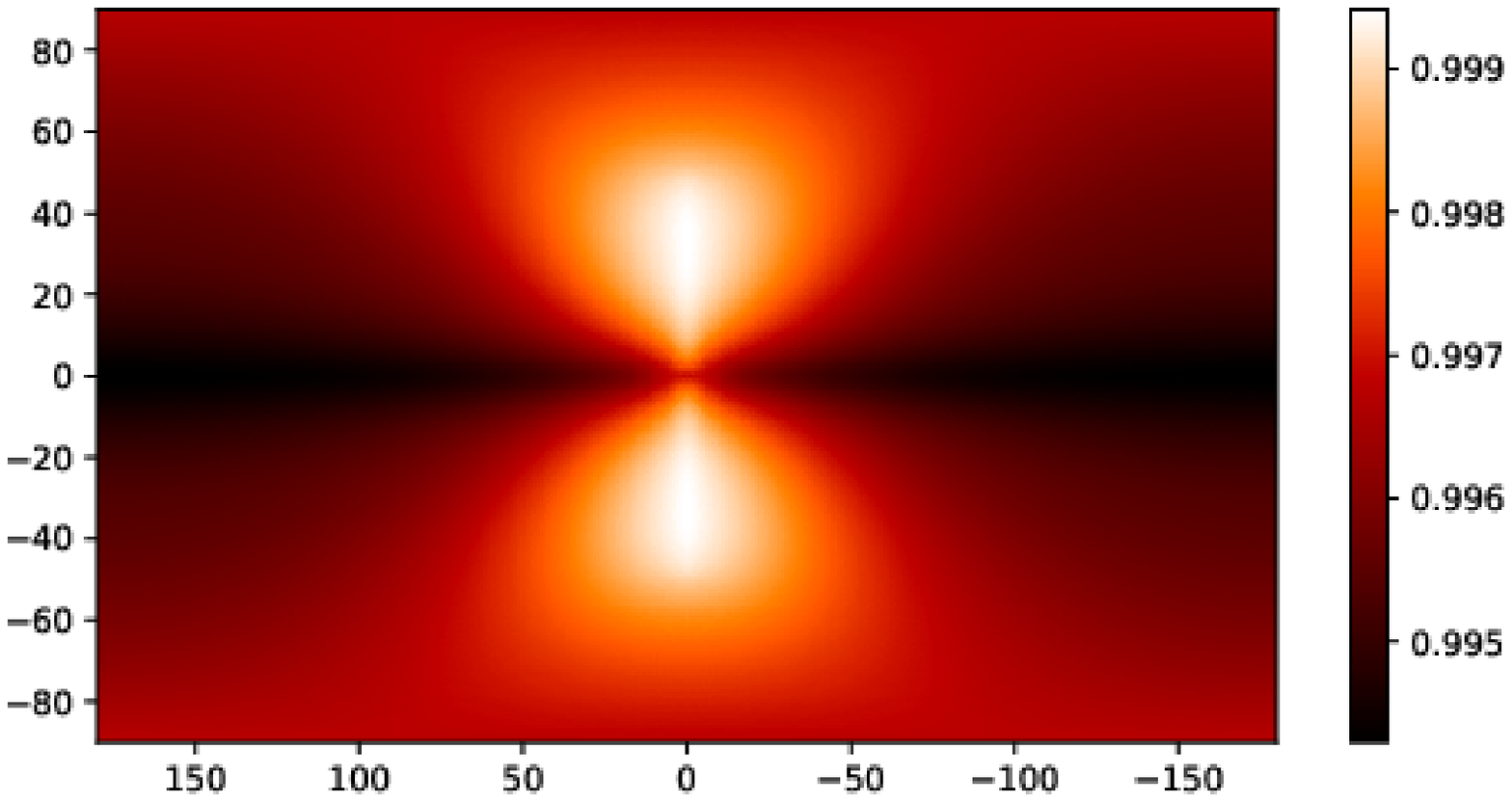}}\\
    \subfloat[Pion decay]{\includegraphics[angle=0,width=0.32\textwidth]{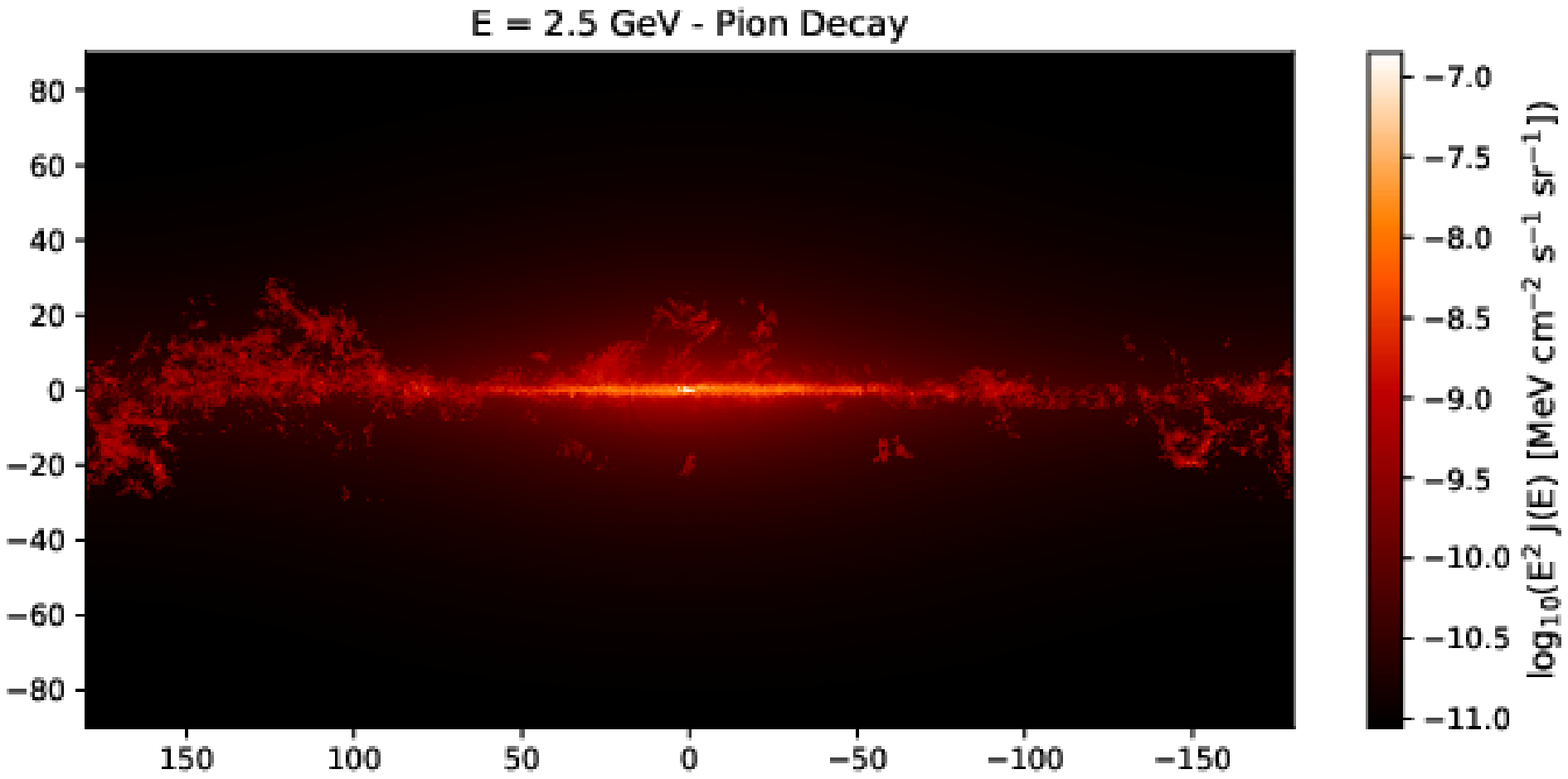}}
    \subfloat[Pion decay]{\includegraphics[angle=0,width=0.32\textwidth]{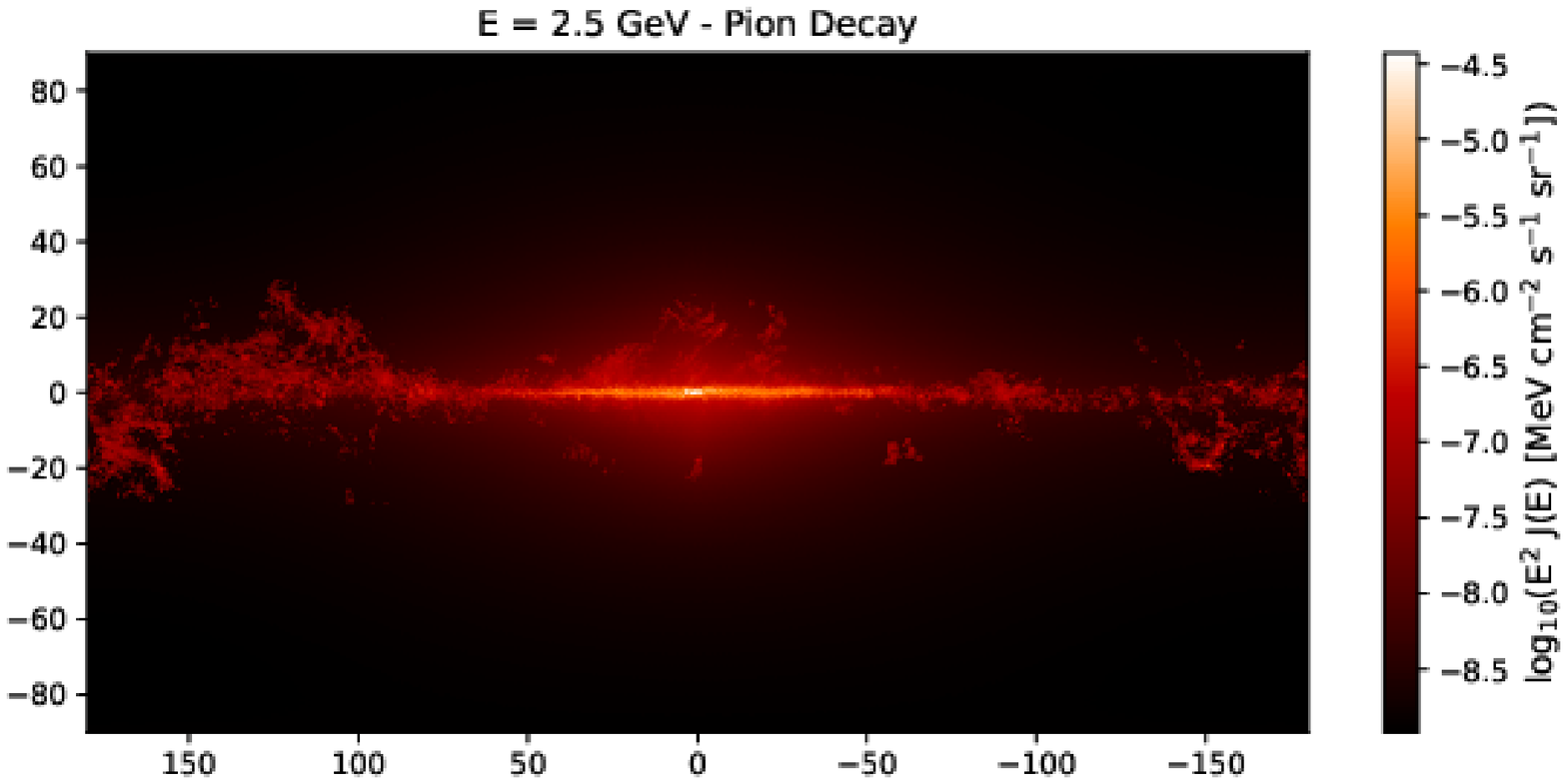}}
    \subfloat[Fractional residual - $\pi$-decay intensity]{\includegraphics[angle=0,width=0.31\textwidth]{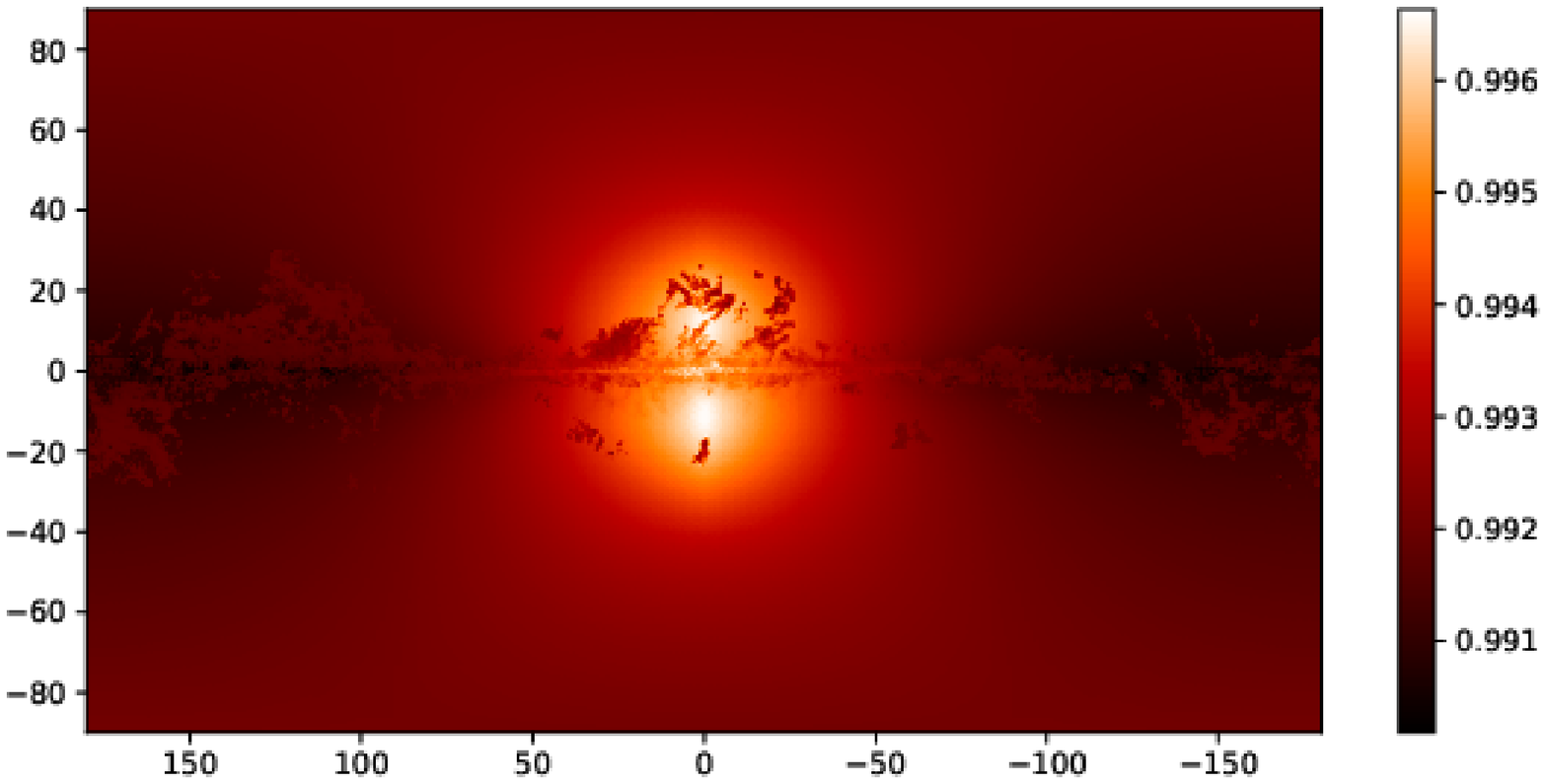}}\\
    \caption{Model of diffuse galactic gamma rays from the Milky Way at 2.5 GeV \cite{fermilat}. Left: Model GC with central proton source. Middle: Model FB with proton source. Right: Fractional residual maps comparing the models. The total gamma-ray emission is composed of hadronic cosmic-ray interactions with gas and leptonic interactions with the gas (bremsstrahlung) and inverse Compton scattering. Maps are in galactic coordinates with $(l,b) = (0,0)$ at the center of the map.}
    \label{diffusefermi}
\end{figure}

\begin{figure}[h!]
  \centering
    \subfloat[Bremss]{\includegraphics[angle=0,width=0.32\textwidth]{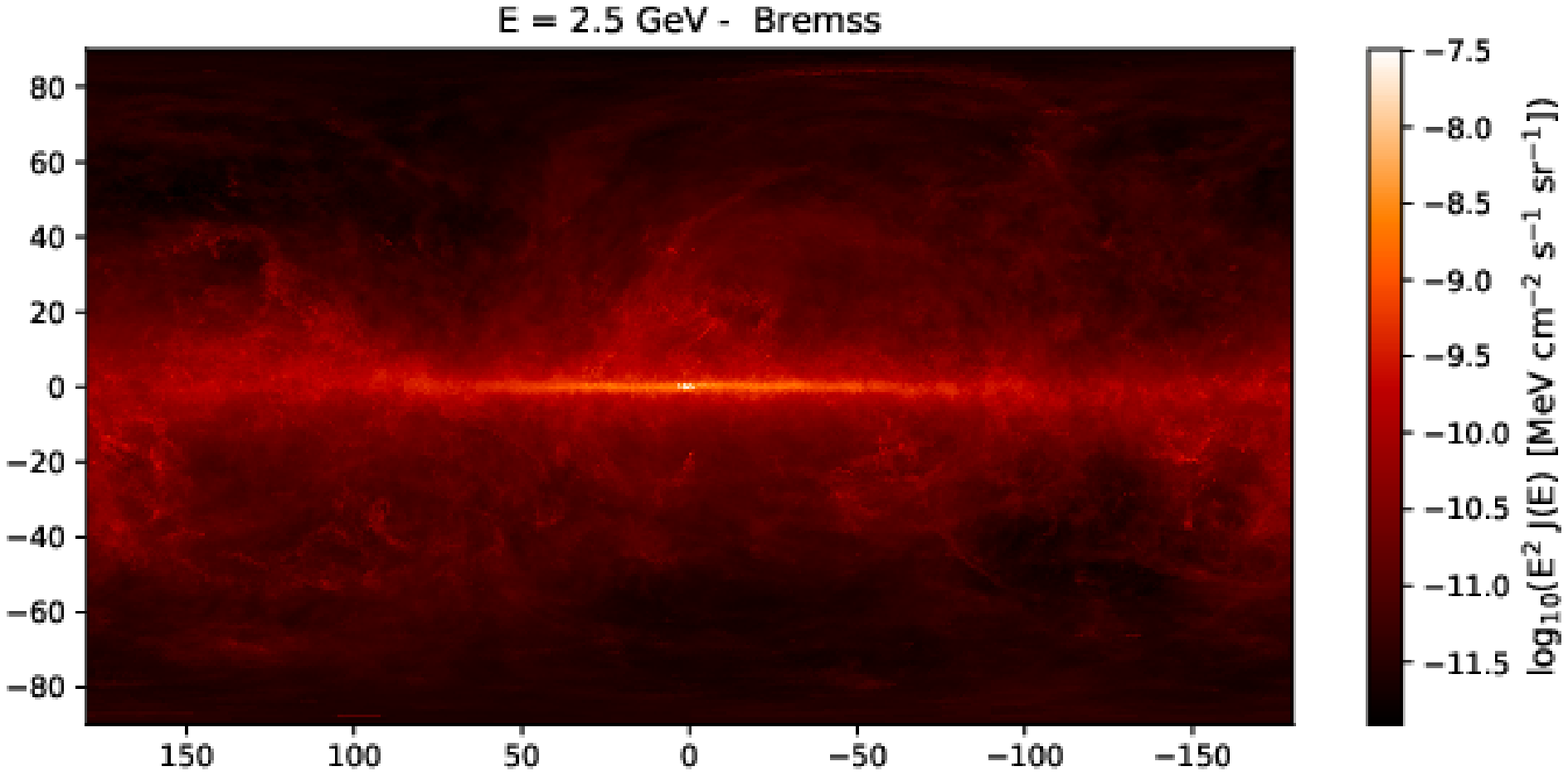}}
    \subfloat[Bremss]{\includegraphics[angle=0,width=0.32\textwidth]{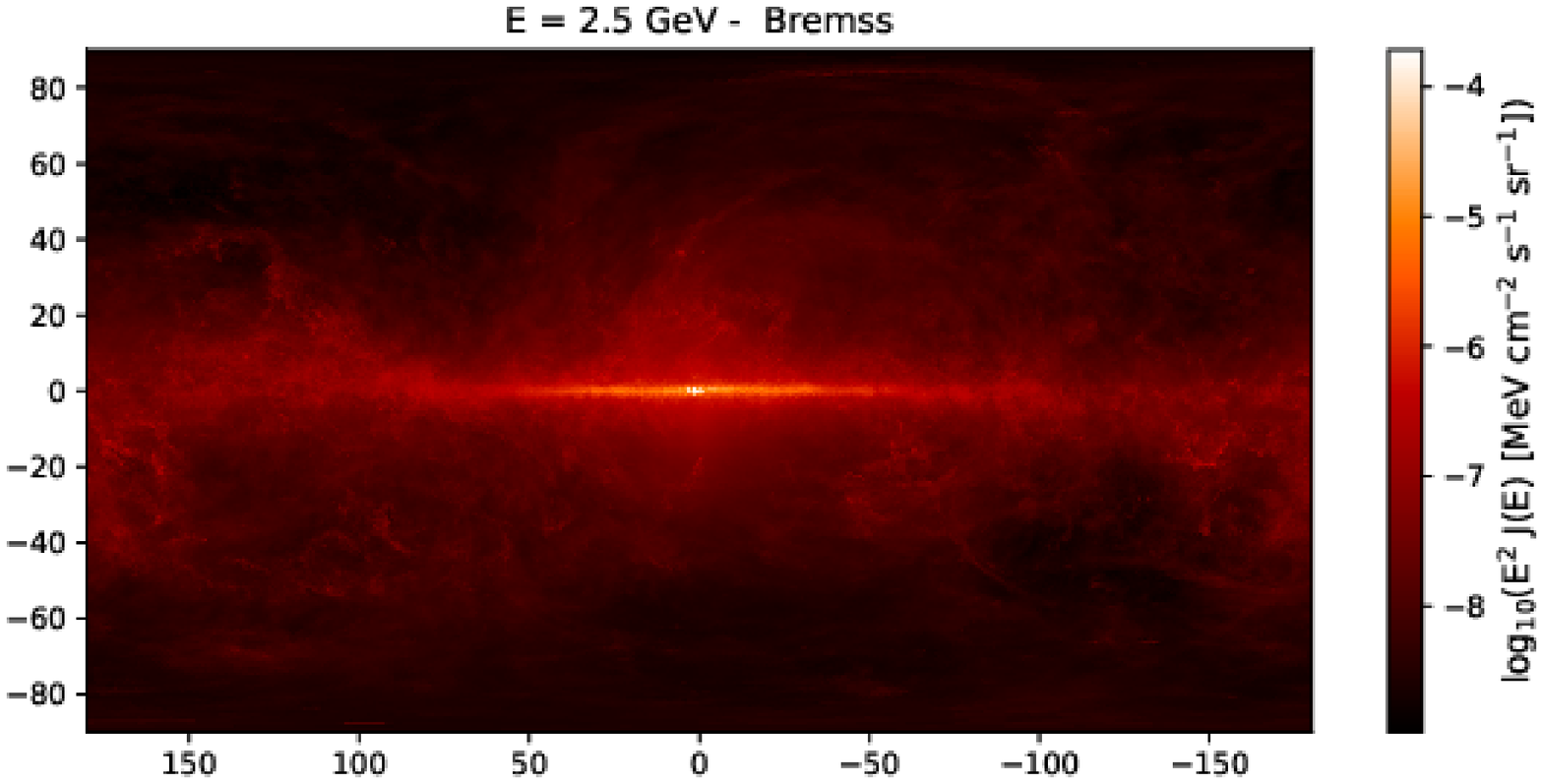}}
    \subfloat[Fractional residual - bremsstrahlung intensity]{\includegraphics[angle=0,width=0.31\textwidth]{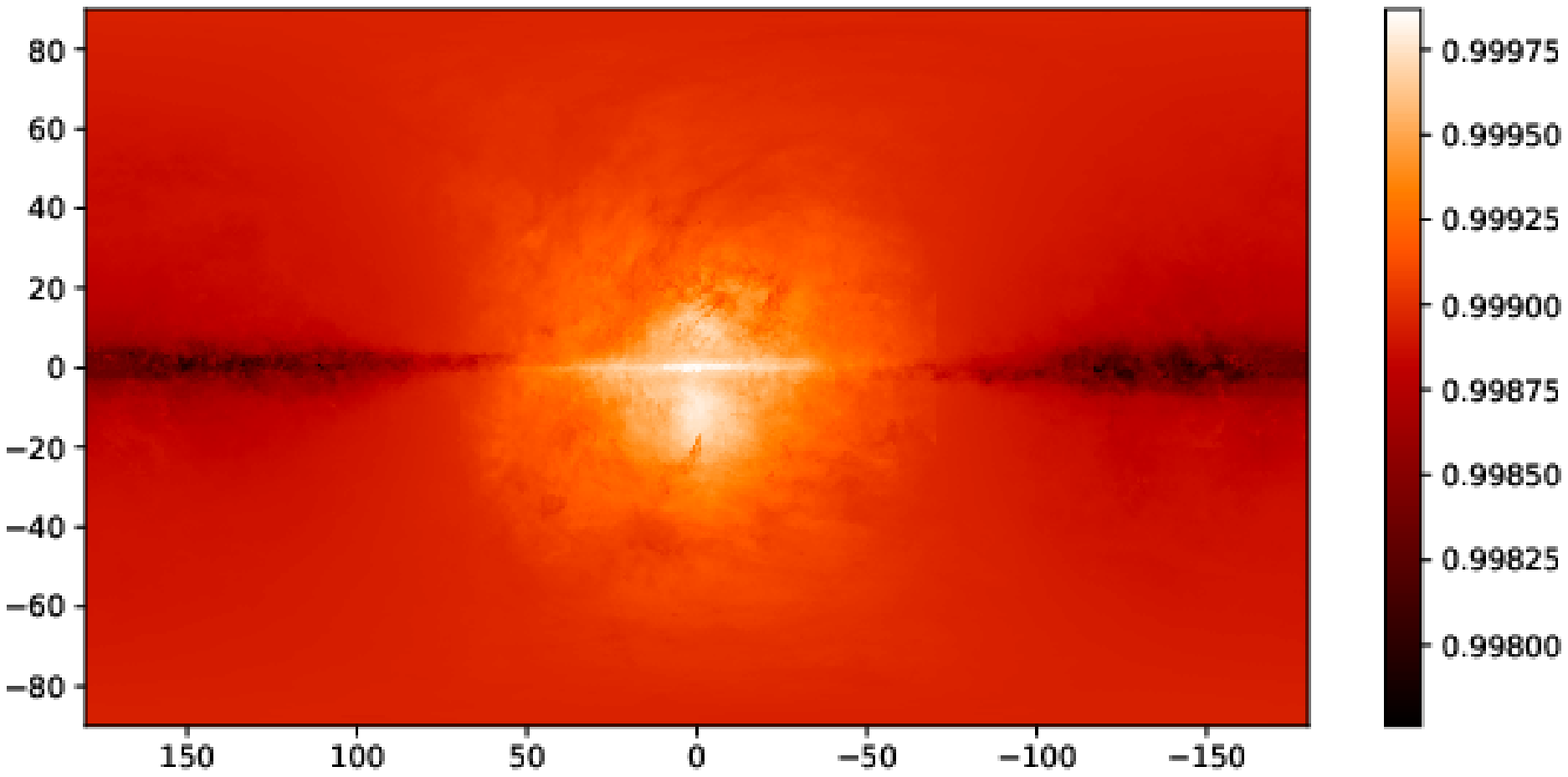}}\\
    \subfloat[Inverse Compton]{\includegraphics[angle=0,width=0.32\textwidth]{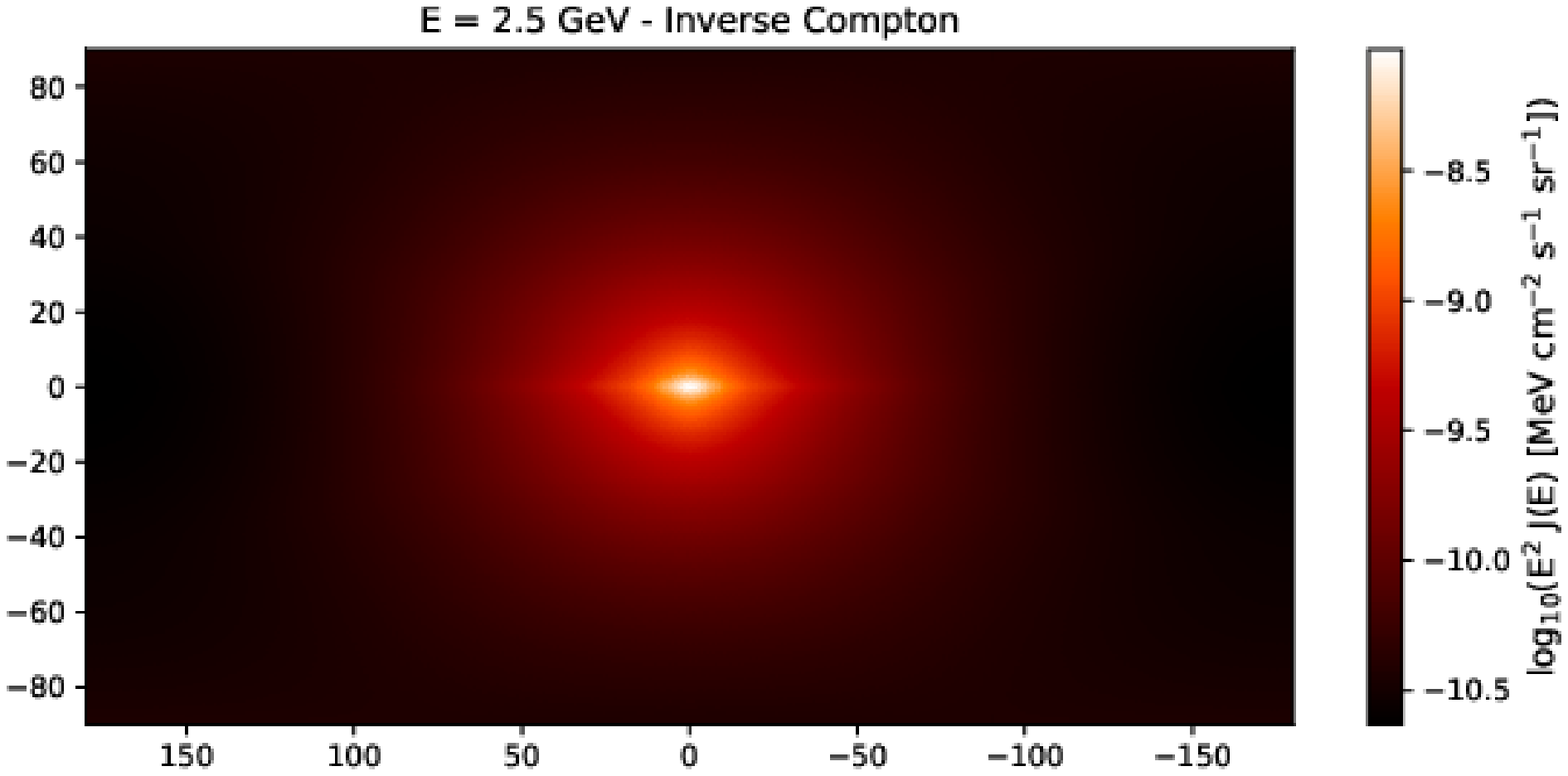}}
    \subfloat[Inverse Compton]{\includegraphics[angle=0,width=0.32\textwidth]{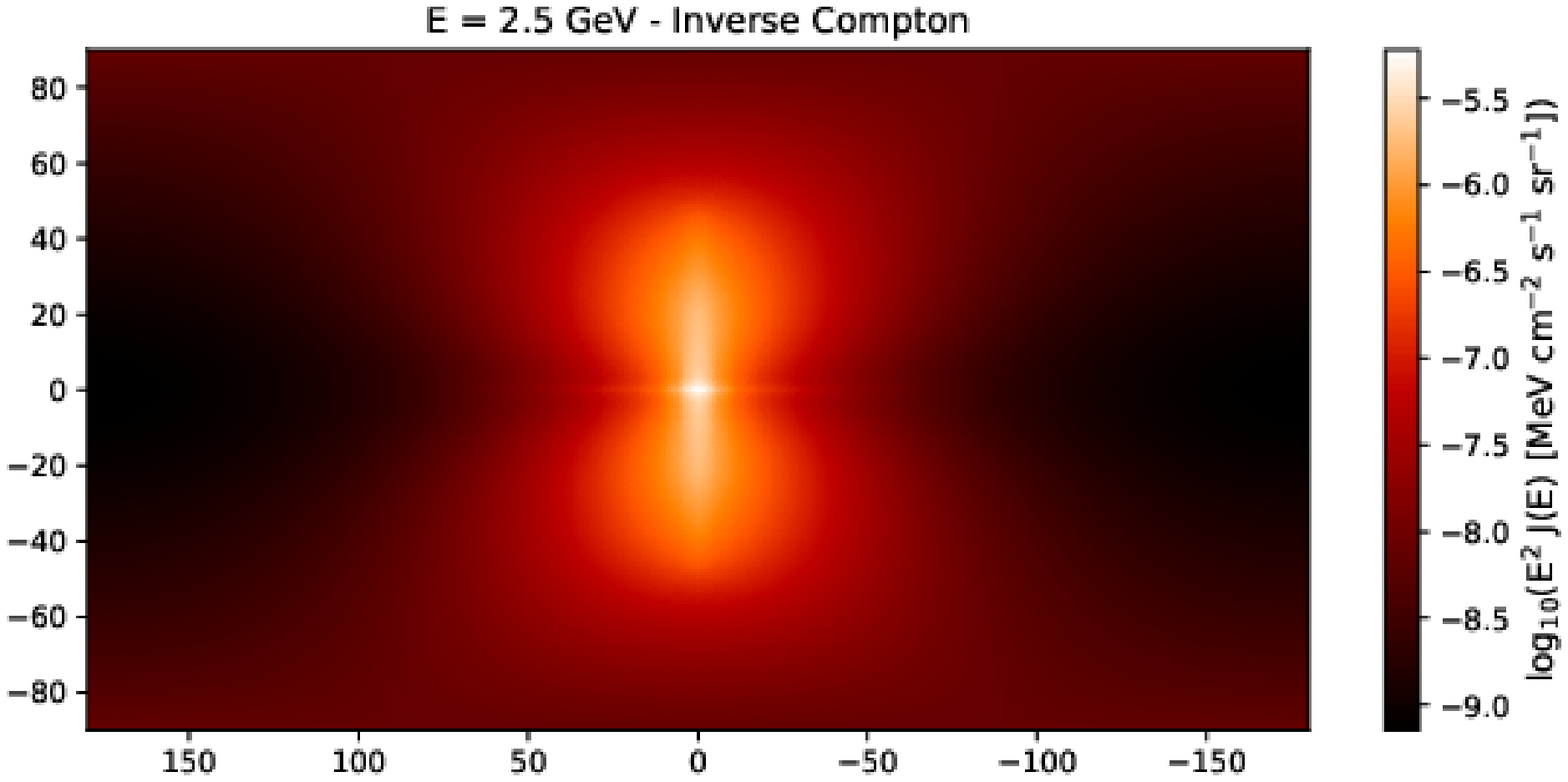}}
    \subfloat[Fractional residual - IC intensity]{\includegraphics[angle=0,width=0.31\textwidth]{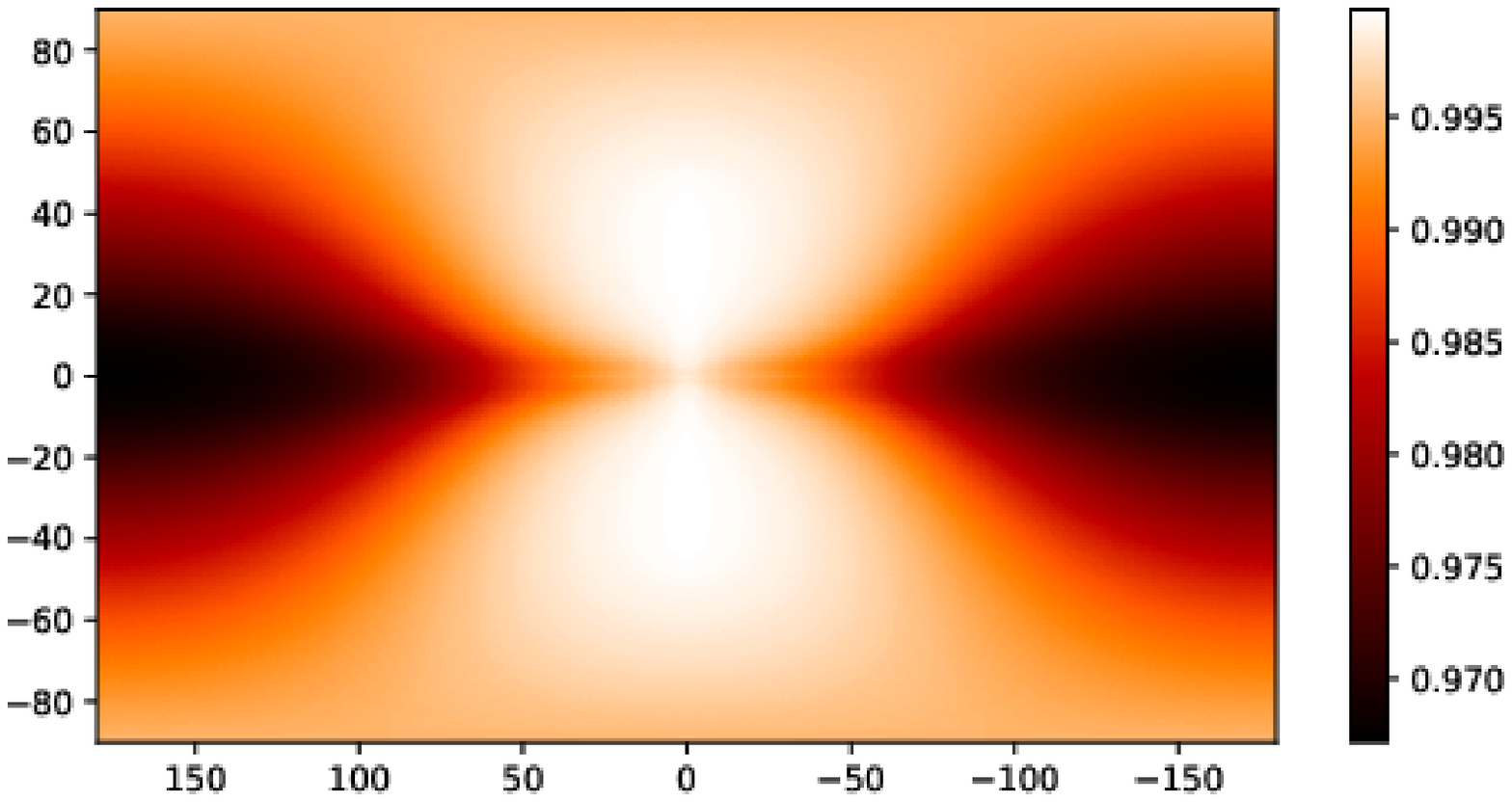}}\\
    \subfloat[Pion decay]{\includegraphics[angle=0,width=0.32\textwidth]{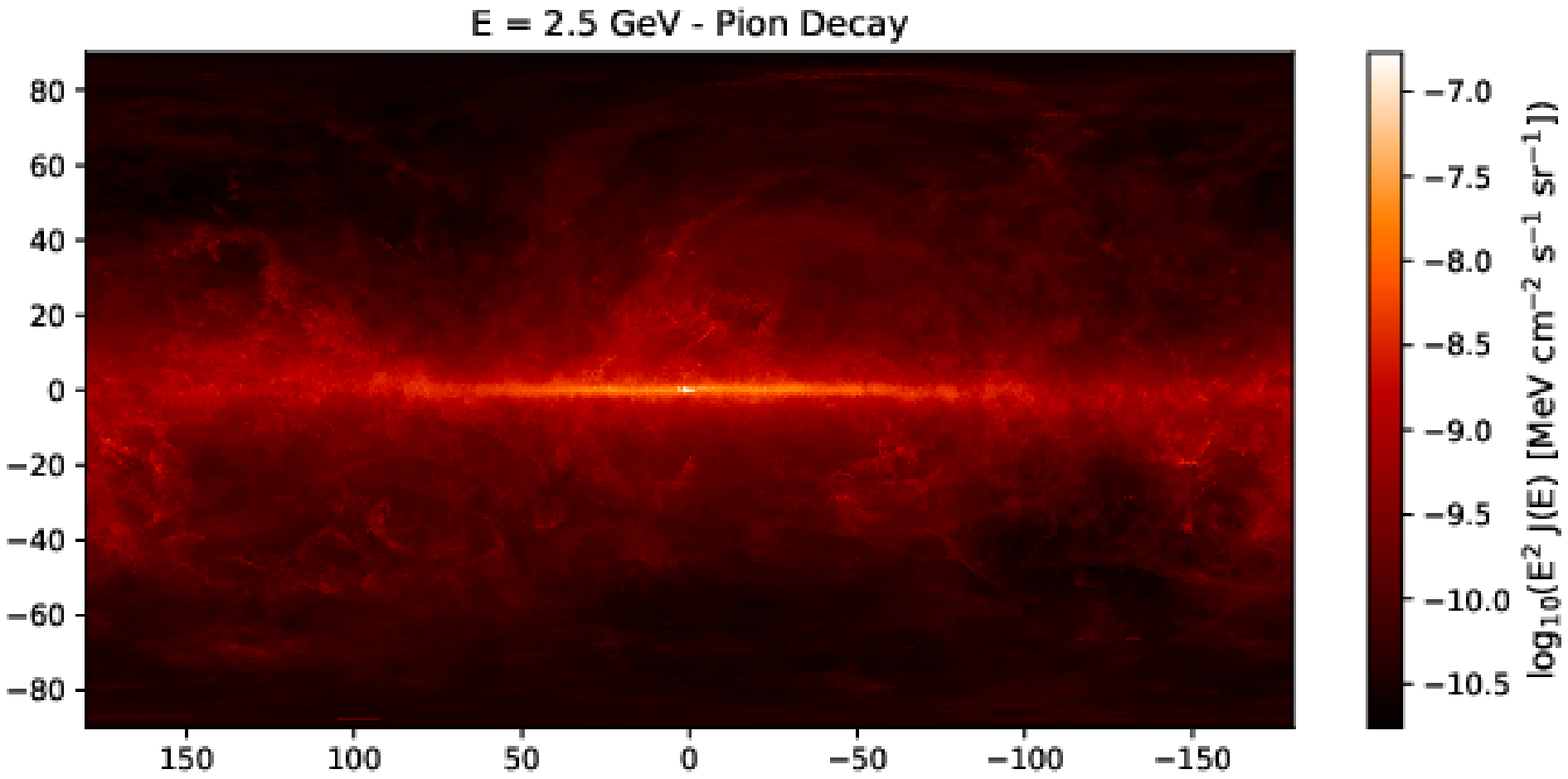}}
    \subfloat[Pion decay]{\includegraphics[angle=0,width=0.32\textwidth]{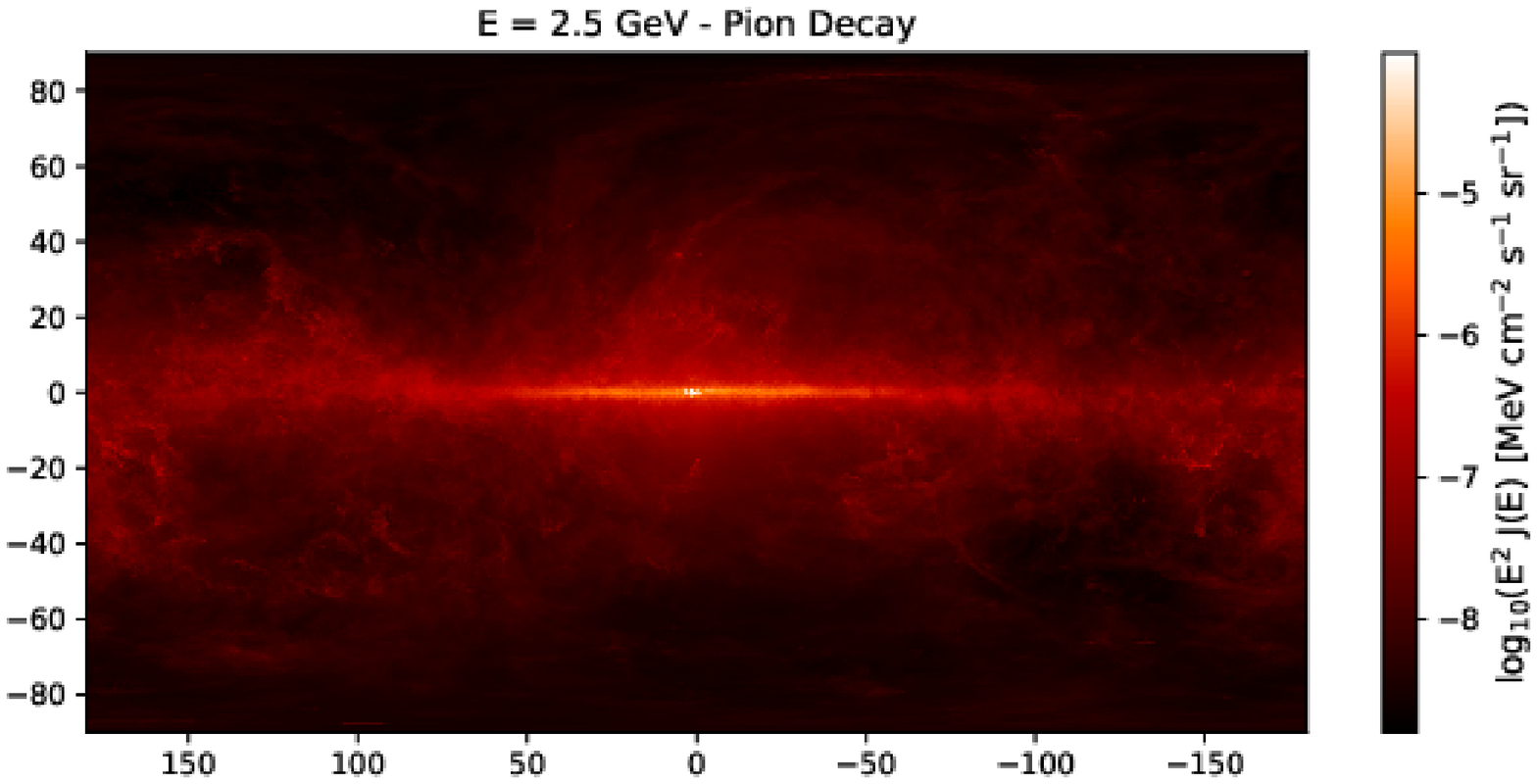}}
    \subfloat[Fractional residual -  $\pi$-decay intensity]{\includegraphics[angle=0,width=0.31\textwidth]{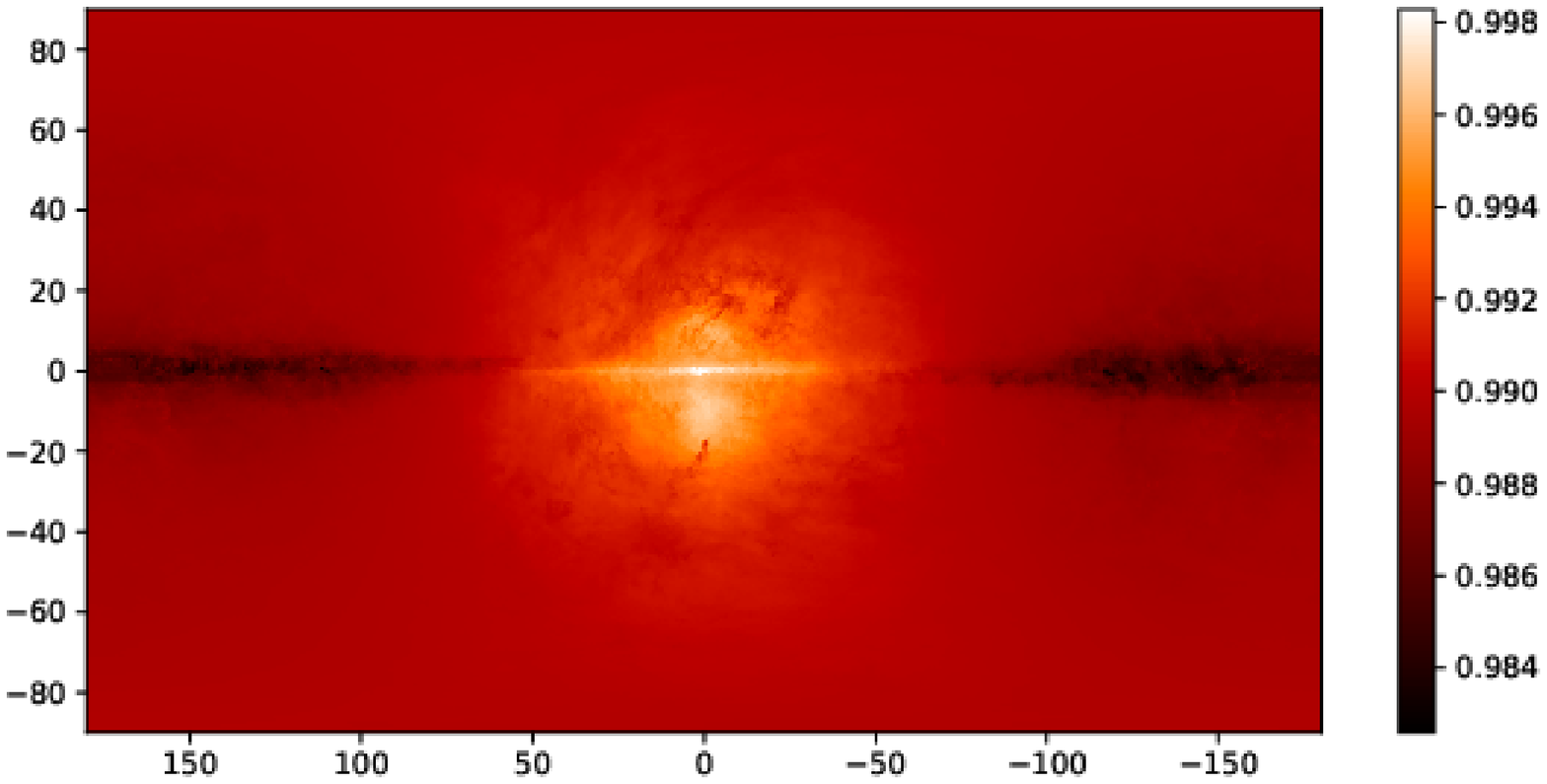}}\\
    \caption{Model of diffuse galactic gamma rays from the Milky Way at 2.5 GeV with 2D gas distribution galactic ensemble component \cite{troy}. Left: Model GC with central proton source. Middle: Model FB with proton source. Right: Fractional residual maps comparing the models. The total gamma-ray emission is composed of hadronic cosmic-ray interactions with gas and leptonic interactions with the gas (bremsstrahlung) and inverse Compton scattering. Maps are in galactic coordinates with $(l,b) = (0,0)$ at the center of the map.}
    \label{diffuse2Dnew}
\end{figure}

The interstellar radiation field (ISRF) is treated by GALPROP considering the contribution of several stellar components and includes the effects of absorption and re-emission from dust in the interstellar medium \cite{web}. The ISRF is divided into three basic components: direct emission from stars, emission from dust grains and CMB. The pion emission is correlated with the distribution of gas in the Galaxy as shown in figures \ref{diffusefermi} - \ref{diffuse2Dnew}. The bremsstrahlung process is also gas correlated, but is lower in intensity $\gamma_{br}/ \gamma_{\pi^{0}}$ $\sim$ 4R, where $R \sim 0.1$ is the ratio of electron to proton CRs. As an example, in figures \ref{diffusefermi} - \ref{diffuse2Dnew}, we show the three different diffuse emission components at 2.5 GeV from our models, which have parameters defined in table 2. The figures \ref{diffusefermi} - (d,e) and \ref{diffuse2Dnew} - (d,e) correspond to the IC emission, which depend primarily on the electron distribution and the properties of the magnetic field and the ISRF. 
Taking into account the properties of the CRs propagation, the figures \ref{diffusefermi} - \ref{diffuse2Dnew} display the leptonic and hadronic interactions with the gas. In order to quantify the impact of the processes intensity between models GC and FB on the spectrum we show the fractional residual maps in figure \ref{diffusefermi} - \ref{diffuse2Dnew} - (c,f,i). It can be observed a structure related to the Bubbles at GC. In Ref. \cite{eric1} was shown that the galactic diffuse emission can affect the nature and the very existence of the GC excess. In addition, the gas models affect the calculations of CRs propagation and the modeling of the gamma-ray emission. The effect of the alternative gas maps on the GC spectrum is shown in figure \ref{diffuse2Dnew}. The intensity at 2.5 GeV for the three components of the interstellar emission shows the effects due to the gas density combination. These maps show that the gas models can influence significantly the bremsstrahlung and pion decay emissions in the inner galaxy, as can be seen in figures \ref{diffuse2Dnew} - (a,b,g,h) and \ref{gamma} - (a,b). This implies that the gas distribution is an important factor to describe the propagation of CRs and the gamma ray emission.

\begin{table}[h!]
\centering
\caption{Summary of Galactic parameters for our models. See text for details.}
\label{tab:1}
\begin{tabular}{c c c c}\hline
Parameters & Units & Model GC  & Model FB \\ \hline
  Source & $-$ & GC & Fermi Bubbles \\
   $D_0$                         & $cm^{2}s^{-1}$   &  $2.7\times 10^{28}$ &  $2.0\times 10^{28}$ \\ 
   $\delta$    & $-$ &  0.6   &     0.6 \\
   L    & $kpc$ &  10.0   &   10.0\\    
   z    & $kpc$ &  20.0   &   20.0\\ 
   $v_{a}$   & $km\;s^{-1}$ &  28.0  &   28.0\\  
   $dv/dz$   & $km\;s^{-1}\;kpc^{-1}$ &  10.0  &   10.0\\          
   $\alpha$   & $-$ &  2.4  &   2.4\\ \hline        
 \end{tabular}
\end{table}

\section{CONCLUSIONS}

There is a general consensus that the origin and acceleration of the galactic cosmic rays of energy up to the knee are related to the distribution of supernova remnants in the galactic disk. However, with the increasing number of studies of high energy phenomena at the Galactic Centre it is becoming clear that this region can not be neglected.

The aim of this work was to study the contribution of the Galactic Centre as a source of galactic CRs based on the GC gamma-ray observations. We have shown that a single source at the GC can describe the low-energy cosmic rays spectra and the observed cosmic-ray ratio $B/C$. We have also carried out simulations of the emission processes of gamma rays using a new 2D galactic distribution gas \cite{troy}, demonstrating the necessity to include the contribution of the gas in these processes.

The results here obtained can be extended to scenarios beyond the knee allowing different contributions of the SNR. For instance, a two-component SNR model: one old SNR component, which would dominate the flux below $\sim$ 100 GeV and the galactic SNR ensemble, which would provide the high-energy flux up to the knee \cite{nicola}. These models are able to describe well the data up to PeV and the spectral shape of the $B/C$ ratio. Another possibility is to apply our description in the two-halo model: CRs propagation taking place in two regions characterized by different energy dependencies of the diffusion coefficient \cite{feng}.

The simulations presented in this paper enable a better interpretation of CRs and gamma rays data from GC and FB, demonstrating that the GC cosmic rays can contribute significantly to the CRs spectrum.  In addition, mechanisms of CRs acceleration at supermassive black hole Sagittarius A* and millisecond pulsars located around the GC can contribute as Pevatrons to the galactic CRs around the knee \cite{fujita, guepina}. We expect that, in the near future, neutrino and TeV gamma-ray detectors, such as CTA \cite{cta}, with its improved resolution and sensitivity, will be able to reveal the very nature of the Galactic Centre gamma ray source.

\begin{acknowledgments}

The  research  of  RCA  is  supported  by  the  CAPES-HARVARD  Program  (Junior Visiting Professor Award) under grant 88881.162283/2017-01, CNPq grant numbers 307750/2017-5 and 401634/2018-3, and Serrapilheira Institute grant number Serra-1708-15022. RCA also thanks to Avi Loeb for very fruitful discussions about the Galactic Centre and the Harvard-Smithsonian Center for Astrophysics for hospitality. The authors acknowledge FAPESP Project 2015/15897-1 and also gratefully the National Laboratory for Scientific Computing (LNCC/MCTI, Brazil) for providing HPC resources of the SDumont supercomputer, which have contributed to the research results reported within this paper.  URL:http://sdumont.lncc.br. The authors thank Daniel Supanitsky for helpful comments and the use of the software package GALPROP v56 \cite{web}.

\end{acknowledgments}

\end{document}